%% file: main.tex
\documentclass[a4paper,11pt]{article}
\usepackage{jcappub}

\usepackage[table,svgnames,dvipsnames]{xcolor}
\usepackage[modulo,mathlines]{lineno}
\usepackage[normalem]{ulem}
\usepackage{aas_macros}
\usepackage{subfigure}
\usepackage{graphicx} 
\usepackage{graphics}
\usepackage{dcolumn} 
\usepackage{bm} 
\usepackage{cases} 
\usepackage{booktabs}
\usepackage{comment}
\usepackage{multirow}
\usepackage{makecell}
\usepackage{siunitx}
\usepackage{tabularx}
\usepackage{xspace}
\usepackage{soul} 
\graphicspath{{figures/}}
\usepackage{fontawesome}
\usepackage{hyperref} 
\hypersetup{colorlinks=true,citecolor=blue,filecolor=blue,urlcolor=blue,linkcolor=blue}
\usepackage{adjustbox}
\usepackage{orcidlink} 
\usepackage{hanging} 
\usepackage{amsmath,mathtools}

\def\be{\begin{equation}}
\def\ee{\end{equation}}

\def\ba#1\ea{\begin{align*}#1\end{align*}}

\renewcommand{\emph}[1]{\textit{#1}}

\usepackage[nameinlink,noabbrev]{cleveref}
\crefname{equation}{Eq.}{Eqs.}
\crefname{section}{Section}{Sections}
\crefname{figure}{Figure}{Figures}
\crefname{table}{Table}{Tables}
\crefname{appendix}{Appendix}{Appendices}
\Crefname{figure}{Figure}{Figures}
\Crefname{equation}{Equation}{Equations}
\Crefname{section}{Section}{Sections}
\Crefname{table}{Table}{Tables}
\crefformat{equation}{\eqA eq.~\eqB #2#1#3)}
\newcommand{\eqA}{}
\newcommand{\eqB}{(}

\newcommand{\mksym}[1]{\ifmmode {\rm #1}\else #1\fi}

\begin{document}

\title{DESI DR2 reference mocks: clustering results from \textsc{Uchuu}-BGS and LRG}

\input{DESI-2024-0514_Mar24}
\emailAdd{efdez@iaa.es}


\abstract{
The aim of this work is to construct mock galaxy catalogues that accurately reproduce the redshift evolution of galaxy number density, clustering statistics, and baryonic properties, such as stellar mass for luminous red galaxies (LRGs) and absolute magnitude in the $r$-band for the bright galaxy sample (BGS), based on the first three years of observations from the Dark Energy Spectroscopic Instrument (DESI). To achieve this, we applied the subhalo abundance matching (SHAM) technique to the \textsc{Uchuu} $N$-body simulation, which follows the evolution of 2.1 trillion particles within a volume of 8 $h^{-3}$Gpc$^{3}$, assuming a Planck base-$\Lambda$CDM cosmology. Using SHAM, we populated \textsc{Uchuu} subhalos with LRGs and BGS-BRIGHT ($r<19.5$) galaxies up to redshift $z=1.1$, assigning stellar masses to LRGs and luminosities to BGS galaxies (up to $M_{\rm r}\leq20$). Furthermore, we analyzed the clustering dependence on stellar mass and luminosity for each tracer. Our results show that the \textsc{Uchuu} BGS-BRIGHT and LRG mocks accurately reproduce the observed redshift evolution of clustering, with better than 5$\%$ agreement for separations of $1<r<20\,h^{-1}$Mpc and below a 10$\%$ for $0.1<r<1\,h^{-1}$Mpc. For the \textsc{Uchuu}-LRG mock, we successfully captured the stellar mass dependence of clustering, while for the \textsc{Uchuu}-BGS mock, we replicated the clustering for various volume-limited subsamples. We also find good agreement between the data and mocks in the dependence of large-scale bias on luminosity for BGS-BRIGHT galaxies and on stellar mass for LRGs. Altogether, these results equip DESI with robust tools for generating high-fidelity lightcones for the remainder of the survey, thereby enhancing our understanding of the galaxy–halo connection.}

\maketitle

\tableofcontents

\section{Introduction}
The use of simulated lightcones has become an essential tool in modern cosmology, particularly in the study of large-scale structure (LSS). These simulations aim to replicate the distribution of galaxies and matter in the universe, providing insights into its evolution and underlying physics \citep{2013MNRAS.435..743D, 2014MNRAS.437.2594W, 2016MNRAS.460.1173R, 2017MNRAS.470.4646S, 2019arXiv190102401D, 2023arXiv230606315P, 2024MNRAS.532.1659E, 2024MNRAS.528.7236D}. By generating mock observations that closely resemble real survey data, researchers can validate theoretical models, refine observational strategies, and interpret cosmological measurements.

The concept of simulating cosmic structures traces back to the mid-20th century when early computational efforts sought to understand the dynamics of gravitational clustering. With the advent of large-scale galaxy surveys, such as the Sloan Digital Sky Survey (SDSS) \citep{2021PhRvD.103h3533A} and the Two-Degree Field Galaxy Redshift Survey (2dFGRS) \citep{2001MNRAS.328.1039C}, the need for more realistic and comprehensive mock catalogues became apparent. These early surveys revealed the cosmic web—a vast network of galaxies, clusters, and voids—and posed challenges in modelling the complex interplay of gravitational, hydrodynamical, and astrophysical processes shaping this structure.

As galaxy surveys expanded in scope and precision, the development of simulations advanced in parallel. Numerical cosmological simulations, such as the \textsc{Millennium} \citep{2005Natur.435..629S} and \textsc{Illustris} \citep{2015A&C....13...12N} simulations, have been pivotal in predicting the clustering of dark matter and its connection to galaxy formation. Coupled with methods like the halo occupation distribution (HOD) and subhalo abundance matching (SHAM), these simulations enable the creation of lightcones that mimic observed properties of galaxies, such as luminosity, stellar mass, and clustering statistics.

Subhalo abundance matching (SHAM) \citep{2002ApJ...569..101M, 2004ApJ...609...35K, 2004MNRAS.353..189V, 2006ApJ...647..201C, 2016MNRAS.460.1173R, 2016MNRAS.460.1457S, 2021MNRAS.505..325G, 2022ApJ...940...13D} has emerged as a powerful method for populating dark matter halos with galaxies. By assuming a monotonic relationship between a halo property, such as peak circular velocity or maximum mass (i.e. the peak value of the maximum circular velocity V$_{\rm max}$ over the history of the halo), and a galaxy property, such as stellar mass or luminosity, SHAM allows for a straightforward and physically motivated mapping of galaxies onto halos. This method requires minimal assumptions about the underlying physics of galaxy formation, making it particularly attractive for constructing mock catalogues that match observed galaxy statistics. Compared to HOD methods, which model the probability distribution of galaxies within halos as a function of halo mass, SHAM provides a direct galaxy-halo connection that naturally captures environmental dependencies. Moreover, SHAM-based mocks are particularly well suited for generating weak lensing, peculiar velocity, and next-generation survey mocks.

The \textsc{Uchuu} \citep{ishiyama} reference mocks we present in this work exemplify the strengths of SHAM-based lightcones by providing high-fidelity galaxy distributions tailored for large-scale structure analyses. These mocks serve as a crucial resource for training and testing covariance mocks, modelling the evolution of clustering with redshift, and capturing the dependence of clustering on baryonic properties. In particular, each Luminous Red Galaxy (LRG) in the mock is assigned a stellar mass, while each Bright Galaxy Sample (BGS) galaxy is assigned an $r$-band absolute magnitude and $g-r$ colour. Furthermore, they provide an accurate determination of halo occupation and bias, ensuring robust interpretations of observed galaxy clustering.

The Dark Energy Spectroscopic Instrument (DESI) \citep{2016arXiv161100036D, 2016arXiv161100037D} represents a major milestone in galaxy surveys, offering an unprecedented view of the universe with millions of spectroscopically confirmed galaxies. To maximize the scientific return of the three years of measurements of DESI, it is crucial to have mock catalogues that replicate the survey’s selection functions and statistical properties. Simulated lightcones for specific galaxy populations, such as LRGs and BGS, play a key role in this endeavour. These mocks allow researchers to: 1) test the robustness of cosmological analyses by providing a controlled environment where the true underlying parameters are known; 2) quantify systematic effects, such as survey incompleteness and redshift errors, which can bias results; 3) optimize measurement techniques, including the estimation of the two-point correlation function and higher-order statistics; and 4) support the development of theoretical models by comparing predictions with observational data.

In this work, we present simulated lightcones designed to mimic the clustering statistics of DESI for Luminous Red Galaxies and the Bright Galaxy Survey from the first three years of operations (DESI DR2, hereafter). These mocks are constructed using the SHAM method applied to the \textsc{Uchuu} halo simulation \citep{ishiyama}, one of the largest and most detailed $N$-body simulations available. By incorporating realistic number density evolution with redshift and accurately reproducing the two-point correlation function, among other key observables, our simulations bridge the gap between theory and observation, serving as a critical resource for advancing our understanding of the universe.

This article is structured as follows: in section \ref{desi}, we provide a brief summary of DESI Data Release 1 (DR1) and of the three first years of measurements from DESI, and describe in detail the galaxy populations used in this work. In section \ref{uchuu}, we introduce the high-resolution simulation used: \textsc{Uchuu}. In section \ref{sham}, we present a detailed explanation of the SHAM methodology applied to construct BGS-BRIGHT and LRG mocks, as well as the procedure for generating the lightcone. Then, in section \ref{results}, we discuss the two-point correlation function of the mocks, comparing them with measurements from DESI DR2 and Y1 at small scales and also large scales, and also study the dependence of this function with absolute magnitude for different volume-limited samples for the case of the BGS and with different stellar mass for the case of the LRGs. We also study the redshift-space distortions. In section \ref{power_spectrum} we present the results from the power spectrum, and, finally, in section \ref{hod_bias} we  analyse the Mean Halo Occupancy and large-scale bias. To end this work, in section \ref{summary}, we summarize our key findings.

\section{DESI DR1 and DR2}\label{desi}

The Dark Energy Spectroscopic Instrument (DESI) is a robotic, fiber-fed, highly multiplexed spectroscopic surveyor that operates on the Mayall 4-meter telescope at Kitt Peak National Observatory \citep{2022AJ....164..207D}. DESI, which can obtain simultaneous spectra of almost 5000 objects over a $\sim$ 3º field \citep{2016arXiv161100037D, 2023AJ....165....9S, 2024AJ....168...95M}, is currently conducting a five-year survey of about a third of the sky. This campaign will obtain spectra for approximately 40 million galaxies and quasars \citep{2016arXiv161100036D}.

The goal of DESI is to determine the nature of dark energy through the most precise measurement of the expansion history of the universe ever obtained \citep{2013arXiv1308.0847L}. DESI was designed to meet the definition of a Stage IV dark energy survey with only a 5-year observing campaign. Forecasts for DESI \citep{2016arXiv161100037D} predict a factor of approximately five to ten improvement on the size of the error ellipse of the dark energy equation of state parameters $w_0$ and $w_a$ relative to previous Stage-III experiments \citep{2006APS..APR.G1002A}.

In this work, we use the DESI Year 1 (Y1) and Year 3 (Y3) catalogues for two main reasons. First, stellar mass estimates based on CIGALE are only available for the Y1 Luminous Red Galaxy (LRG) sample (see \cite{2024arXiv240919066S} for a detailed explanation of the CIGALE tool, and subsection \ref{sham-lrg} for a description of its application in this work). Second, the DESI Collaboration has not yet released pre-reconstruction measurements of the two-point correlation function (TPCF) at Baryon Acoustic Oscillation (BAO) scales for Y3. As a result, comparisons at these scales can only be performed using Y1 data. However, at smaller scales, we carry out the comparison using the Y3 TPCF.

\subsection{DESI DR1}
The DESI Data Release 1 (DR1, DESI DR1 hereafter) dataset \citep{desi_dr1} contains observations made with the DESI instrument during its first year of operations. These observations occurred during the main survey operations, which began on May 14, 2021, following a period of survey validation \citep{2024AJ....167...62D}, and continued until June 13, 2022. The DESI instrument captures spectra from 5,000 targets simultaneously by using robotic positioners to place optical fibres at the targets' celestial coordinates \citep{2023AJ....165....9S}. These fibres are grouped into ten "petals", which direct the light to ten separate climate-controlled spectrographs.

DESI data is gathered through observations of "tiles", where each tile corresponds to a specific pointing on the sky with a field of view of approximately 8 square degrees. For each tile, targets are assigned to the 5,000 robotic fibres according to a prioritization scheme. DESI allocates its observation time between "bright" and "dark" time programs based on observing conditions. "Bright time" refers to nights with substantial moonlight or twilight periods, while "dark time" refers to nights with little or no moonlight. This distinction is important because the sky background significantly affects the detection of faint objects. DESI also employs a "backup" program during particularly poor observing conditions.

During DR1, 2,744 tiles were observed during "dark" time, and 2,775 tiles were observed during "bright" time. The bright galaxy sample \cite[BGS]{2023AJ....165..253H} is observed during bright time, while the luminous red galaxies \cite[LRG]{2023AJ....165...58Z}, quasars \cite[QSO]{2023ApJ...944..107C}, and emission line galaxies \cite[ELG]{2023AJ....165..126R} are observed during dark time. This division optimizes the use of telescope time by targeting brighter objects when sky conditions are less favourable and reserving the best observing conditions for fainter, higher-redshift targets. The data were first processed by the DESI spectroscopic pipeline \citep{2023AJ....165..144G} the morning after the observations for immediate quality checks and later underwent a uniform processing run (internally referred to as "iron") to generate the redshift catalogues used in this paper.

The BGS sample consists of two types of targets: BGS-BRIGHT and BGS-FAINT. The BGS-BRIGHT sample includes galaxies with magnitudes $r < 19.5$, while the BGS-FAINT sample covers the range $19.5 < r < 20.175$ and has also a selection in color to achieve high redshift efficiency \citep{2023AJ....165..253H}. The majority of BGS targets fall into the BGS-BRIGHT category, which is the sample used in this work. 

In this work we will focus on the creation of the \textsc{Uchuu}-LRG and \textsc{Uchuu}-BGS lightcones. The methodology used in order to create the \textsc{Uchuu}-ELG and \textsc{Uchuu}-QSO lightcones is presented in \citep{vaisakh}, as well as the main results from the clustering statistics.

\subsection{DESI DR2}
On the other hand, we will utilize the galaxy catalogues containing data from the first three years of operations, spanning from May 14, 2021, to April 2024, which includes DR1 \citep{baoy3_desi, baoy3_desi2}. This catalogue, referred to as DESI DR2, includes 6,671 tiles observed during 'dark' time and 5,171 tiles observed during 'bright' time. These numbers represent 2.4 and 2.3 times the number of tiles released in DR1, respectively.

\begin{table*}
\caption{Basic properties of the DESI BGS-BRIGHT Y1 and Y3 (first and second row, respectively) and LRG Y1 and Y3 (third and fourth row, respectively): the redshift interval, median redshift (z$_{\rm med}$), effective area of the sky footprint weighted by completeness (A$_{\rm eff}$), number of galaxies weighted by completeness (N$_{\rm eff}$), and effective volume (V$_{\rm eff}$).}
    \centering
    \begin{tabular}{c|ccccccc}
    \toprule
         Sample&  Redshift range&  z$_{\rm med}$&  A$_{\rm eff}$ (deg$^{2}$) & N$_{\rm eff}$& V$_{\rm eff}$ ($h^{-3}$Gpc$^{3}$)\\
        \midrule
         Y1 BGS-BRIGHT&  $0.02<z<0.5$& 0.203 & 7473 & 6189576 & 1.73 \\
        Y3 BGS-BRIGHT&  $0.02<z<0.5$& 0.203 & 12416 & 10069613 & 2.88 \\
         Y1 LRG&  $0.4<z<1.1$& 0.753 & 5740 & 3013218 & 7.98 \\
         Y3 LRG&  $0.4<z<1.1$& 0.753 & 10077 & 5172900 & 14.01 \\         
         
        \bottomrule
    \end{tabular}
    \label{summary_targets}
\end{table*}

\begin{figure}
    \centering
    \includegraphics[width=\linewidth]{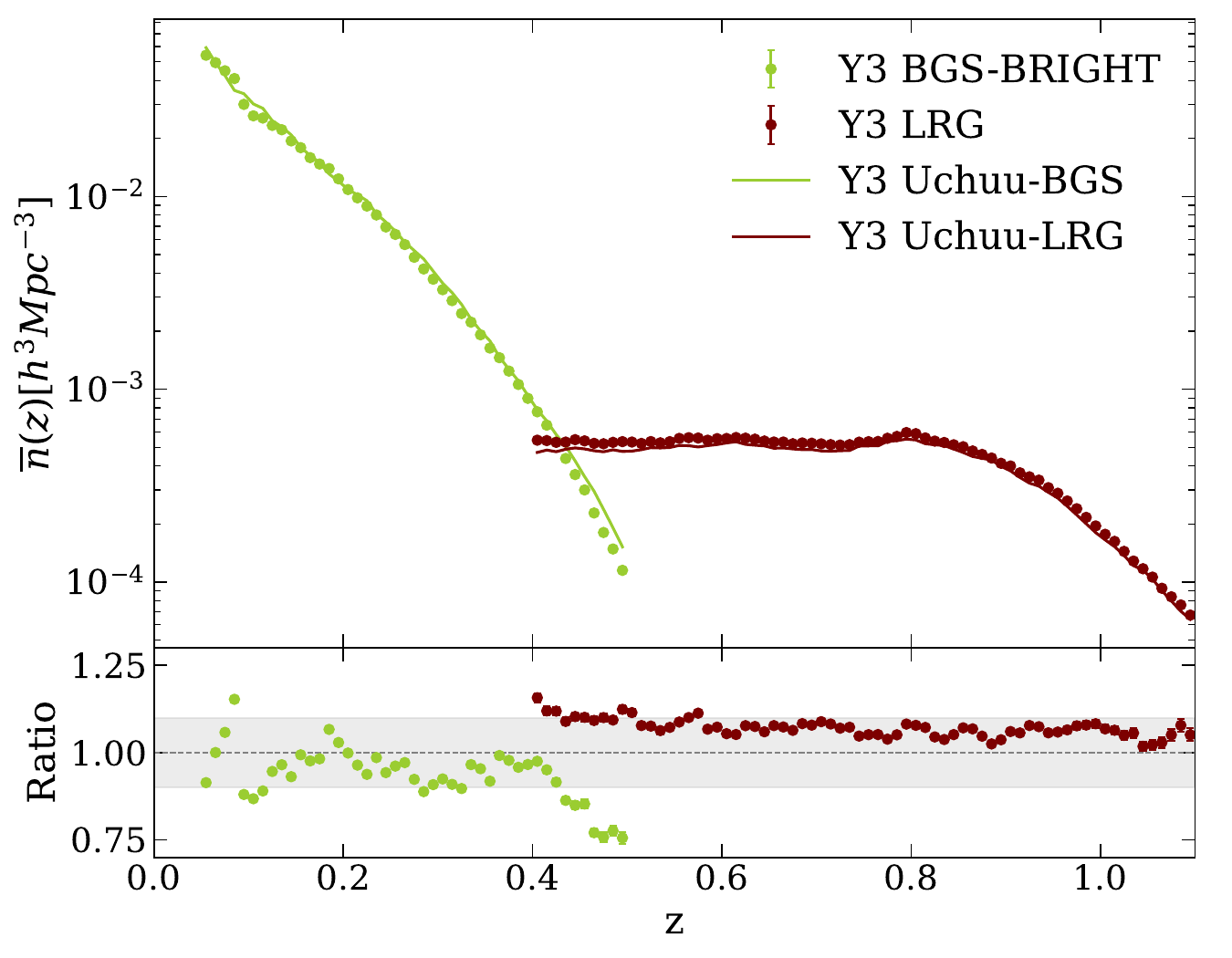}
    \caption{The comoving number density corrected by completeness of Y3 LRG and BGS-BRIGHT DESI (points) and Y3 \textsc{Uchuu} mock lightcones (solid lines) is shown in the top panel. The ratio between data and the prediction of the mock for each target is shown in the bottom panel. The shaded area represents the limits of the 10$\%$.}
    \label{nz}
\end{figure}

In Table \ref{summary_targets} we list the basic properties of the LRG and BGS-BRIGHT samples used in this work. This table includes information such as the redshift ranges, sky area, total weighted number of galaxies and effective volume as a measure of constraining power. The effective area is calculated generating N$_{\rm random}$=1$\times10^{7}$ random points and checking how many of them are inside Y1 or Y3 bright (dark) footprint for the case of the BGS-BRIGHT (LRG), N$_{\rm inside}$. A$_{\rm eff}$ is calculated then as

\begin{equation}
    A_{\rm eff} = \frac{N_{\rm inside}}{N_{\rm random}}\frac{360^{2}}{\pi},
\end{equation}
where $\displaystyle{\frac{360^{2}}{\pi}}$ is the area of the full-sky lightcone in square degrees. V$_{\rm eff}$ is calculated then as

\begin{equation}
    V_{\rm eff} = \frac{4\pi}{3}\frac{N_{\rm inside}}{N_{\rm random}}(r_{\rm max}^{3}-r_{\rm min}^{3}),
\end{equation}
where r$_{\rm min}$ and r$_{\rm max}$ is the comoving distance that corresponds to the maximum and minimum redshift of each target survey (indicated in the second column of table \ref{summary_targets}). We consider a flat $\Lambda$CDM cosmology with Planck-15 parameters \citep{planck2015} when transforming redshift to comoving distances: $h$=0.6774, $\Omega_{\rm m}$=0.3089, $\Omega_{\rm b}$=0.0486, n$_{s}$=0.9667, $\Omega_{\Lambda}$=0.6911, and $\sigma_{8}$=0.8159.

The top panel of Figure \ref{nz} shows the comoving number density for the two Y3 samples. The bottom panel presents the ratio between the data and the mock prediction. It is observed that for redshifts in the range of $0.1 < z < 0.4$, the mock prediction is quite accurate, with the ratio remaining below 10$\%$. However, for higher redshifts ($z > 0.4$), the deviation increases to approximately 25$\%$. For the LRGs, the mock slightly underestimates the number density compared to the data, although the ratio remains below 10$\%$, except near $z \sim 0.4$.

\begin{figure}
    \centering
    \includegraphics[width=\linewidth]{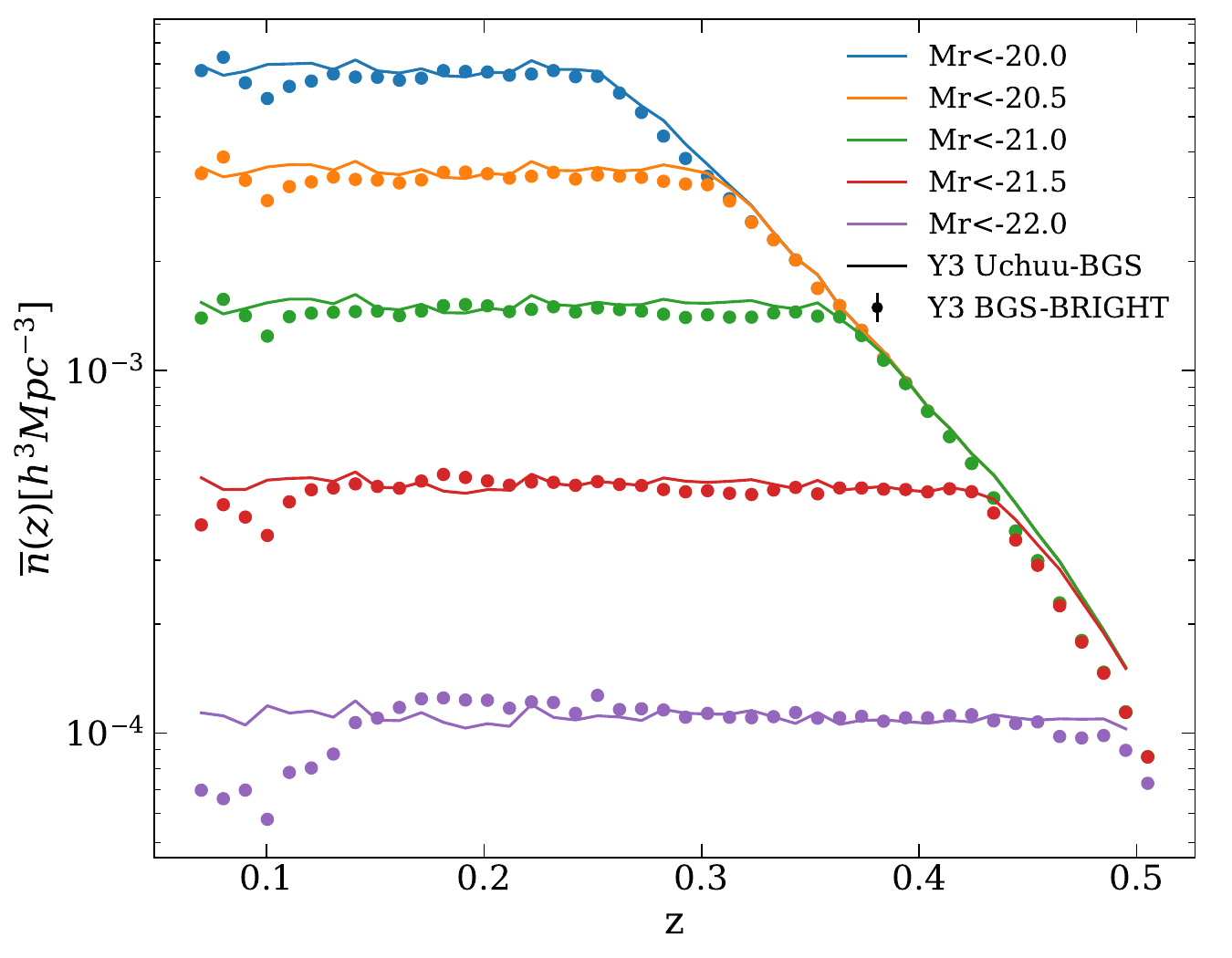}
    \caption{Number density as a function of redshift for different volume-limited samples (detailed in table \ref{bgs_volsam_tab} calculated from the \textsc{Uchuu}-BGS lightcone (solid lines) and DESI BGS-BRIGHT Y3 (points).}
    \label{nz_BGS_volsam}
\end{figure}

In Figure \ref{nz_BGS_volsam} the BGS number density as a function of redshift can be seen for the different volume-limited samples constructed in this work (see table \ref{bgs_volsam_tab} for details). It can be proved that the mean ratio between the data and the mock is about a 5$\%$ for all the volume-limited samples if we take into account only the redshift range where $n(z)$ is approximately constant.

\begin{table}
\caption{Minimum and maximum redshift for different volume-limited samples of the BGS survey, as well as the effective number of galaxies ($N_{\rm eff}$) and the effective volume of each sample.}
    \centering
    \begin{tabular}{ccccc}
    \toprule
       $M_{r}$  & $z_{\rm min}$ & $z_{\rm max}$  & $N_{\rm eff}$  & $V_{\rm eff}$ ($h^{-3}$Gpc$^{3}$)\\
       \midrule
         -22.0 & 0.02 & 0.50 & 311406 & 2.89 \\
         -21.5 & 0.02 & 0.42 & 852877 & 1.83 \\
         -21.0 & 0.02 & 0.34 & 1457497 & 1.03\\
         -20.5 & 0.02 & 0.28 &  1983958 & 0.60 \\
         -20.0 & 0.02 & 0.23 &  2120063 & 0.35\\
         \bottomrule
    \end{tabular}
    \label{bgs_volsam_tab}
\end{table}

\section{The \textsc{Uchuu} simulation}\label{uchuu}
In order to model the clustering signal of the DESI DR1 survey within the flat $\Lambda$CDM Planck cosmology \citep{planck2018}, we utilized the \textsc{Uchuu} $N$-body simulation \citep{ishiyama}. Designed specifically to model the DESI survey, \textsc{Uchuu} features high numerical resolution, which enables the resolution of low-mass dark matter haloes and subhaloes within its large volume. This capability allows us to apply the SHAM technique to populate \textsc{Uchuu} haloes and subhaloes with DESI galaxies, generating mock lightcones to reproduce the number density and predict the clustering of each DESI tracer.

The \textsc{Uchuu} simulation was run using the TreePM code GREEM \citep{art4}. The box has a comoving side length of 2 $h^{-1}$Gpc, with 12,800$^{3}$ dark matter particles. The mass resolution and gravitational softening length are 3.27$\times$10$^{8}$ $h^{-1}$M$_{\odot}$ and 4.27 $h^{-1}$kpc, respectively. The initial conditions were generated using the second-order Lagrangian Perturbation Theory (2LPT) approximation at $z_{\rm init}$=127, and the simulation followed the growth of cosmic structures in the Planck 2015 flat $\Lambda$CDM cosmology \citep{planck2015}. We saved 50 snapshots of the particle distribution from $z$=14 to $z$=0, and identified bound structures using the Rockstar\footnote{The minimum number of particles per halo identified by Rockstar is 10, corresponding to the mass of 3.27e9 Msun/h. The force resolution used in Rockstar is the same with the softening length, and the other parameters were set to the default values.
} phase-space halo$\backslash$subhalo finder \citep{2013ApJ...762..109B}. We constructed merger trees for these structures using a parallel version of the Consistent-Trees algorithm \citep{2013ApJ...763...18B}. Additionally, we obtained the peak value of the maximum circular velocity, V$_{\rm max}$=max$\left(\sqrt{\frac{GM(r)}{r}}\right)$, over the history of each (sub)halo, denoted as V$_{\rm peak}$. We measure the maximum circular velocity at each of the 50 redshift outputs, and take the maximum value as V$_{\rm peak}$. We used this to implement the SHAM method for populating \textsc{Uchuu}  haloes with DESI galaxies. For more information of the simulation methodology and performance, we refer the reader to \cite{ishiyama}. All \textsc{Uchuu} data products are publicly available through Skies $\&$ Universes.\footnote{\url{https://www.skiesanduniverses.org/Simulations/Uchuu}}

\section{SHAM method for BGS-BRIGHT and LRG}\label{sham}
In this section we explain the SHAM method used for BGS-BRIGHT and LRG tracers to assign absolute magnitudes and stellar masses, respectively. The choice of different target properties reflects the distinct nature of these galaxy populations: BGS-BRIGHT is a flux-limited sample where luminosity is the primary observational constraint, while LRG is a color-selected sample where stellar mass better characterizes the underlying galaxy population and its evolution.

\subsection{\textsc{Uchuu} BGS-BRIGHT}
\subsubsection{SHAM Procedure and Target Property Choice}
We follow the SHAM methodology from \cite{2024MNRAS.528.7236D} to construct the \textsc{Uchuu} Y3 BGS-BRIGHT lightcones because: 

\begin{enumerate}
    \item BGS-BRIGHT is a magnitude-limited survey with a flux cut at r=19.5;
    \item luminosity directly relates to the observational selection function;
    \item and the magnitude-limited nature makes absolute magnitude the most physically motivated property for abundance matching.
\end{enumerate}

This approach has previously been applied to BOSS \citep{2016MNRAS.460.1457S}, SDSS \citep{2024MNRAS.528.7236D} and the DESI Early Data Release \citep{2024AJ....168...58D, 2023arXiv230606315P}. SHAM assumes that the most massive (sub)halos host the most luminous galaxies. To generate our simulated flux-limited BGS-BRIGHT sample, we use the luminosity function derived from Y1 DESI BGS-BRIGHT galaxies \cite[see][for more details]{sam_lf}. However, we only consider the luminosity function derived from the North Galactic Cap (NGC), as the one derived from the South Galactic Cap (SGC) systematically underpredicts the observed number density of BGS-BRIGHT galaxies in that region. This discrepancy likely originates from uncertainties in the luminosity computation, such as differences in photometric calibration or $k+E$ corrections between the two regions.

This luminosity function is $k+E$ corrected, allowing us to apply the same function consistently across all \textsc{Uchuu} simulation boxes, each corresponding to a different redshift. The $k$-corrections are computed using polynomial fits based on the FastSpecFit $k$-correction catalogue from DESI \citep{2023ascl.soft08005M}, with a reference redshift of $z_{\rm ref}=0.1$. These corrections transform absolute magnitudes into the rest-frame SDSS $r$-band system, $z=0.1$, ensuring an homogeneous comparison of galaxies across different redshifts. Additionally, we correct for passive evolution using the model of \cite{2014MNRAS.445.2125M}, where the evolution correction $E(z)$ is parametrized as $E(z) = -Q (z - z_{\rm ref})$, with $Q$ determined by minimizing the $V/V_{\rm max}$ distribution to ensure an optimal fit to the observed luminosity evolution, where  V$_{\rm max}$ is the maximum volume over which the galaxy could be seen inside the volume of the survey up to the observed redshift of the galaxy.

\begin{figure*}
    \centering
    \includegraphics[width=0.49\linewidth]{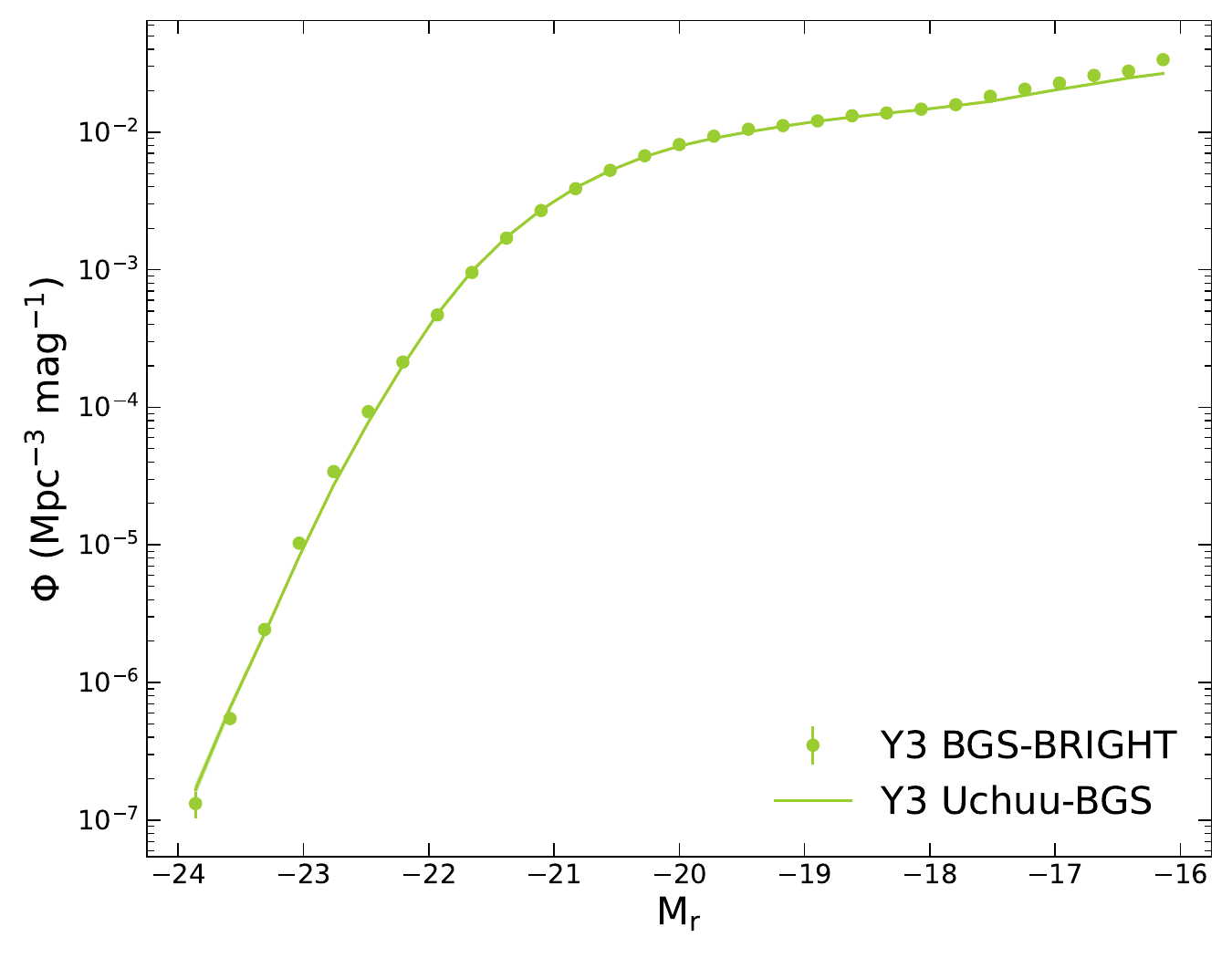}
    \includegraphics[width=0.49\linewidth]{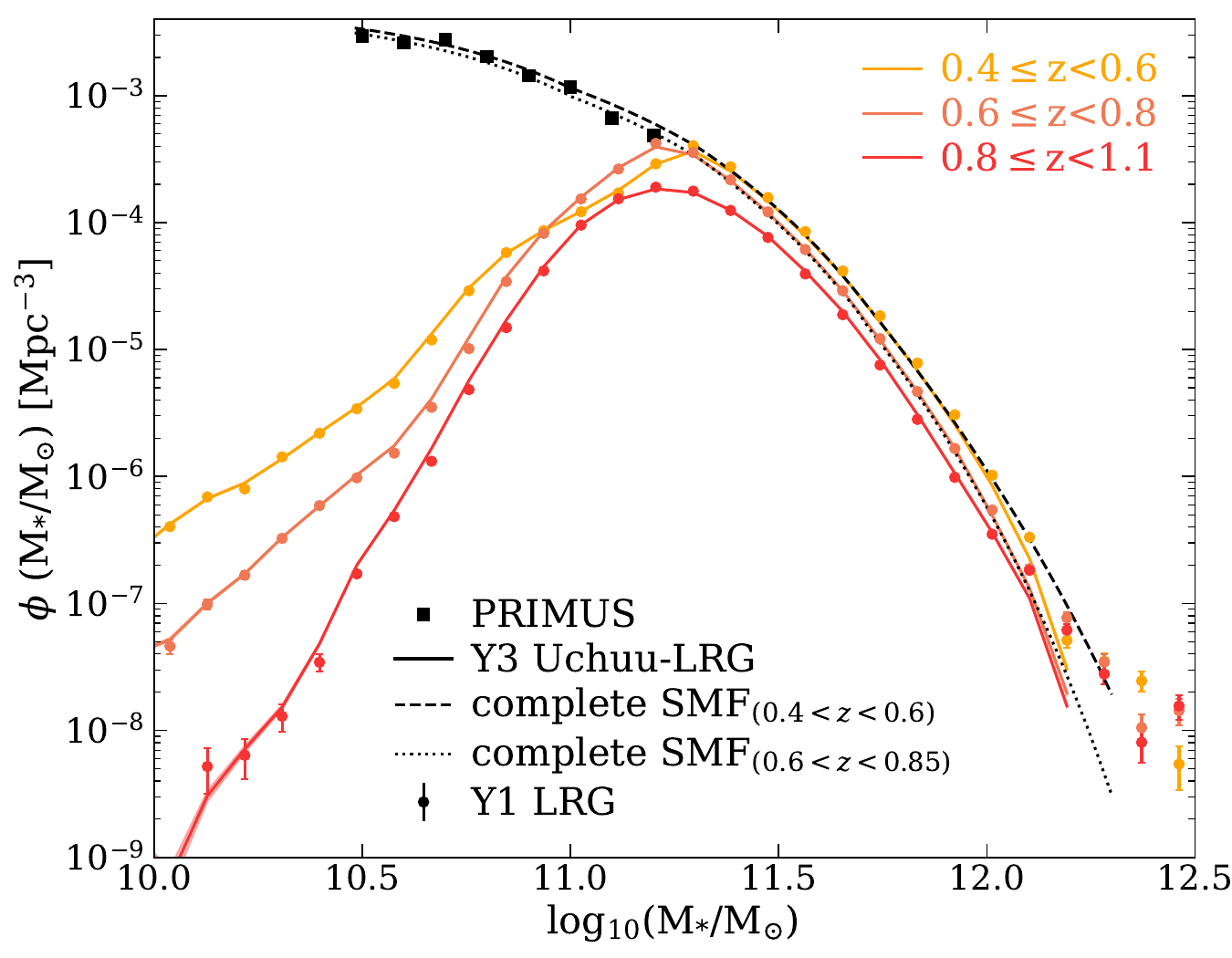}
    \caption{Left: Luminosity function from Y3 NGC BGS-BRIGHT (green points) and \textsc{Uchuu}-BGS (green line). The shaded area and the error bars indicates the 1$\sigma$ uncertainty. Right: stellar mass function of Y1 LRG DESI (points) and Y1 and Y3  \textsc{Uchuu}-LRG lightcones (dotted and solid green lines) are shown for several redshift bins within the range $0.4<z<1.1$. The dashed and dotted black lines represent the complete SMF adopted in each redshift range, indicated in the legend.}
    \label{LF_SMF}
\end{figure*}

The luminosity functions from Y3 BGS-BRIGHT and \textsc{Uchuu}-BGS are shown in the left panel of Figure \ref{LF_SMF}. Uchuu-BGS luminosity function is compatible within 1$\sigma$ with Y3 BGS-BRIGHT luminosity function.

\subsubsection{Step-by-Step SHAM Implementation}
A summary of the SHAM method for BGS is the following:

\begin{enumerate}
    \item Sort the haloes by V$_{\rm peak}$ in decreasing order, to compute the cumulative number density function $n_{\rm h}(>V_{\rm peak})$.
    \item Assign an 'unscattered' galaxy $r$-band magnitude, $\prescript{0.1}{}M^{r}_{h}$, to each halo by matching the above V$_{\rm peak}$ cumulative halo number density function to the target galaxy luminosity function, preserving the ranking such that the large V$_{\rm peak}$ haloes host high-luminosity galaxies:
    \begin{equation}
    n_{g}(< \prescript{0.1}{}M^{r}_{h}) = n_{h}(>V_{\rm peak}).
\end{equation}
\item Define a new 'scattered' value of the absolute magnitude, $\prescript{0.1}{}M^{r, scat}_{h}$:
\begin{equation}
    \prescript{0.1}{}M^{r, scat}_{h} = \mathcal{N}(0, \sigma^{2}) + \prescript{0.1}{}M^{r}_{h},
    \label{scatter_BGS_eq}
\end{equation}
where  $\mathcal{N}(0, \sigma^{2})$ is a normal distribution with median equal to zero and standard deviation $\sigma$, which is the only free parameter of this model.
\item Sort the haloes by $\prescript{0.1}{}M^{r, scat}_{h}$, and compute the $\prescript{0.1}{}M^{r, scat}_{h}$ cumulative distribution, $n_{\rm h}(<\prescript{0.1}{}M^{r, scat}_{h})$.
\item Assign the final $r$-band magnitude by matching the cumulative distribution of $\prescript{0.1}{}M^{r, scat}_{h}$ to the target cumulative distribution function,
    \begin{equation}
    n_{g}(< \prescript{0.1}{}M^{r}_{h}) = n_{h}(>\prescript{0.1}{}M^{r, scat}_{h}).
\end{equation}
\end{enumerate}

The scatter implementation preserves the target luminosity function by design. After introducing scatter in step 3, we perform a re-ranking and re-assignment in step 4 that explicitly matches the scattered halo distribution back to the original target luminosity function. This ensures that the final galaxy population exactly reproduces the input luminosity function, while the scatter affects only the specific halo-galaxy assignments, not the marginal luminosity distribution.

We have verified this preservation by comparing the resulting luminosity function from \textsc{Uchuu}-BGS with the input Y3 BGS-BRIGHT luminosity function (left panel of Figure \ref{LF_SMF}), showing excellent agreement within 1$\sigma$ uncertainties.

Also, it is important to mention that fitting the scatter so that we match the monopole of the 2PCF at small scales\footnote{It can be checked that changing the value of the scatter only affects the 2PCF of to scales of 20-30 Mpc$h^{–1}$} may mask the different cosmology between the data and the mocks. If the real cosmology of the data were approximately the cosmology from Planck15 or slightly different, then the scatter will absorb this difference. However, if the real cosmology were too different from Planck18, then one would expect that the shape of the 2PCF would be different (the peak of the BAO would be shifted, for example, and the scatter has no effect at these scales), and then the scatter wouldn't be enough in order to reproduce the 2PCF from the data.

\subsubsection{Luminosity-Dependent Scatter}
In this work we have found that $\sigma$ is a function of the luminosity. The value of the scatter is 0.05 for galaxies with $\prescript{0.1}{}M^{r}_{h}>-20.5$, and 0.4 for $\prescript{0.1}{}M^{r}_{h}\leq -20.5$. This dependence of the scatter with the absolute magnitude has been calibrated matching \textsc{Uchuu}-BGS TPCF with the one from DESI BGS-BRIGHT. However, we have tested it only for galaxies brighter than $\prescript{0.1}{}M^{r}_{h}<-20.0$. For future works, we will improve the dependence of the scatter with the absolute magnitude so that we can find a functional form, and also reach the faintest galaxies.  However, the dependence of the scatter with luminosity obtained in this work is the one expected from previous works, such as \cite{2019MNRAS.488.3143B}. 

For clarity and readability, we will omit the superscript $0.1$ and the subscript $h$ in $\prescript{0.1}{}M^{r}_{h}$ from this point onward. The superscript $0.1$ indicates that the k-corrections were computed using a reference redshift of $z_{\rm ref} = 0.1$, while the subscript $h$ denotes that the absolute magnitudes have been corrected for evolution.

\subsubsection{Flux-Limited Sample Construction}
The last important step is to remove all those galaxies with apparent $r$-band magnitude fainter than $r=19.5$. We have to do this because, as mentioned above, DESI BGS-BRIGHT is magnitude-limited. In order to do this, we need to transform absolute magnitudes to apparent magnitudes, and for that we need to assign a colour for each galaxy (in order to remove the $k+E$ corrections). Following the methodology described in \cite{2022MNRAS.516.1062S}, 
the rest-frame $g - r$ colour distribution is modelled as a double-Gaussian function, with parameters for the mean and dispersion of both the red and blue sequences, as well as the fraction of blue galaxies. Unlike previous implementations based on SDSS data, this approach employs a smoothly broken linear function to more accurately represent the colour distribution across different magnitudes. These colour distributions are evolved with redshift using the convention of \cite{2014MNRAS.445.2125M} to ensure consistency with DESI BGS-BRIGHT Y1 observations. A more detailed explanation of how the evolution correction is calculated can be seen in \cite{sam_lf}. Additionally, the modelling of colour differences between central and satellite galaxies is refined, improving the overall agreement with observed clustering properties. This enhancement leads to a more realistic colour-dependent clustering behaviour, better aligning the mock catalogue with empirical measurements from DESI BGS-BRIGHT Y1.

However, caution must be taken when using these colours. In Appendix \ref{modelling_colour}, we present a comparison of the probability density functions for the colour distribution in the mock data, the observational data, and the adopted model. This comparison, shown in Figure \ref{color_comparison}, is performed across different bins of magnitude and redshift. By construction, the agreement between the mock data and the model is excellent. However, discrepancies emerge when comparing the model to observational data, especially for the faintest galaxies and in the tails of the distribution (i.e., the reddest and bluest galaxies). These mismatches arise from the model's assumption of a double Gaussian fit, which does not always provide a good approximation in all magnitude–redshift bins. Moreover, while the current methodology allows some flexibility in assigning colours differently to central and satellite galaxies, it does not account for the possibility of central galaxy assembly bias. Investigating this limitation will be an important step toward improving the realism of future mock catalogues.

The sub(halo) boxes we use to construct the \textsc{Uchuu}-BGS lightcone are those with redshift 0.49, 0.43, 0.36, 0.19, 0.093 and 0. As already mentioned above, the luminosity function considered in this work is $k+E$ corrected. This allows us to apply the same function consistently across all the snapshots.

\begin{figure*}
    \centering
        \includegraphics[width=0.49\linewidth]{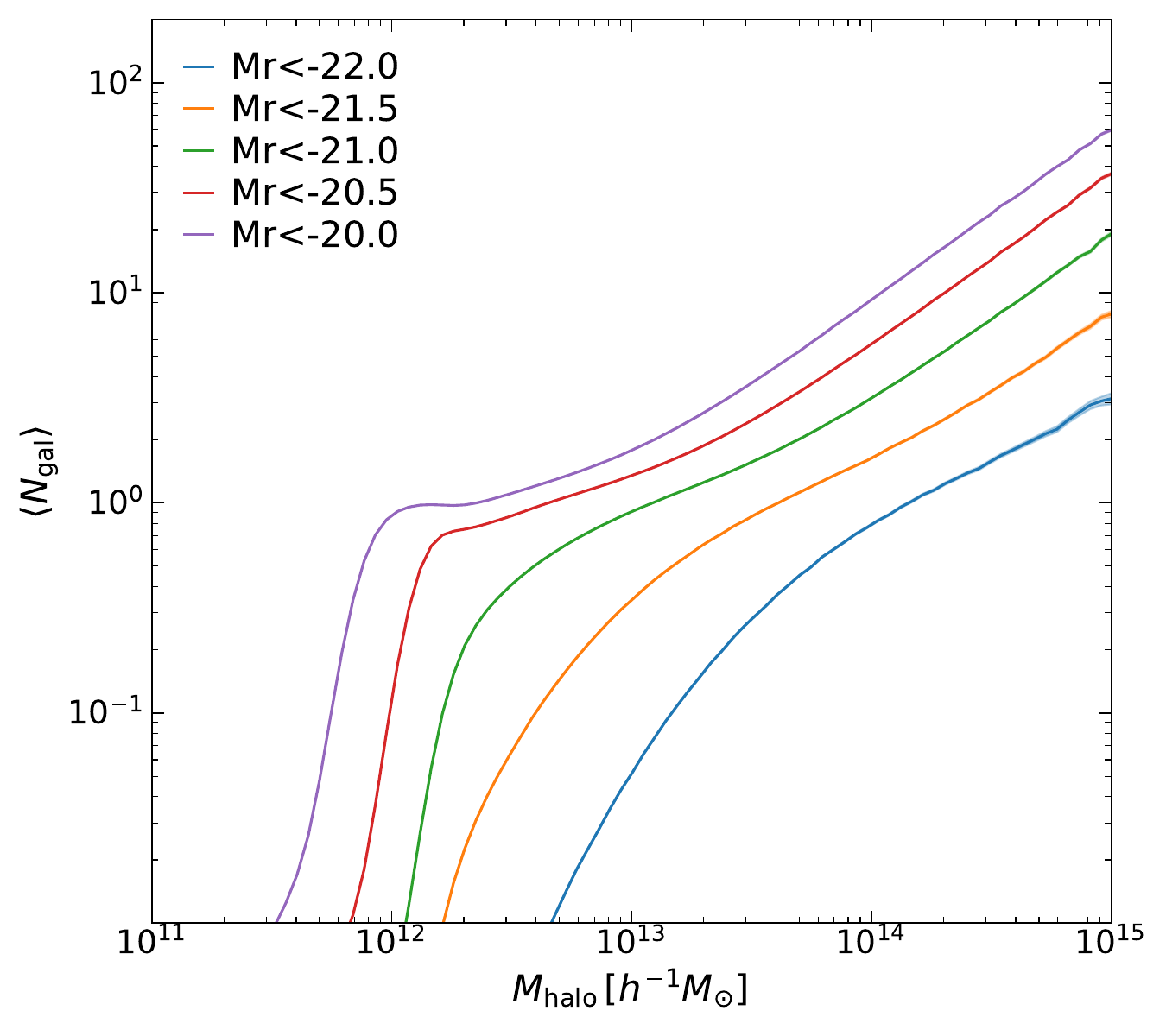}
    \includegraphics[width=0.49\linewidth]{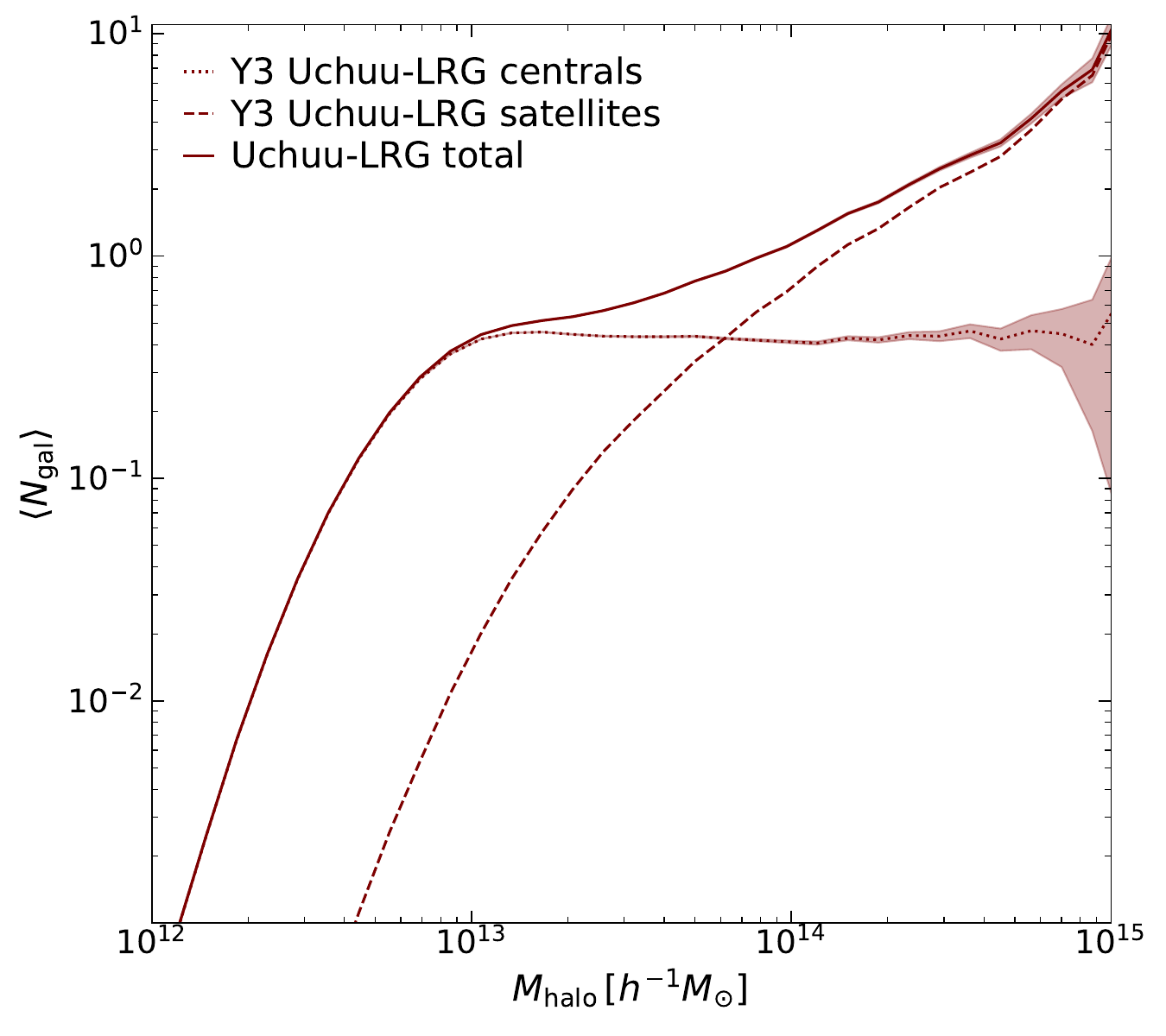}
    \caption{Mean halo occupancy of BGS volume-limited samples (left panel) and LRG samples with stellar mass cuts (right panel) as determined from our SHAM \textsc{Uchuu} boxes. The redshift of the \textsc{Uchuu}-BGS box is $z=0.19$, while the redshift of the \textsc{Uchuu}-LRG box is $z=0.78$. The mean number of galaxies of a halo with a given mass $M_{\rm halo}$ is denoted by $<N_{\rm gal}>$. The solid lines represent the combined centrals and satellite occupation. The shaded area indicates the 1$\sigma$ uncertainty of the occupation measured from the \textsc{Uchuu} lightcone.}
    \label{HOD}
\end{figure*}

In the left panel from Figure \ref{HOD} the halo occupancy distribution (HOD) from the \textsc{Uchuu}-BGS for different volume-limited samples constructed from the galaxy box with $z=0.19$ is shown. We compare the DR2 \textsc{Uchuu}-BGS HOD with that presented in \cite{2023arXiv230606315P}. For the brightest volume-limited sample, the agreement between the two HODs is excellent, while larger discrepancies are observed for the fainter volume-limited samples. These differences are likely driven by a combination of effects. In particular, the relatively small volume of the One-Percent Survey makes its HOD measurements more susceptible to cosmic variance, especially for satellite galaxies. In addition, methodological differences may also play a role: while \cite{2023arXiv230606315P} applies a single scatter in $V_{\rm peak}$ to the full sample, in this work we allow the scatter to vary with magnitude, which can naturally impact the inferred HODs.

\subsection{\textsc{Uchuu} LRG}\label{sham-lrg}
\subsubsection{SHAM Procedure and Target Property Choice}

For LRG, we use stellar mass as the target property because the LRG sample is designed to be complete only for massive, evolved galaxies. Using the luminosity function would result in a severely incomplete distribution across the full luminosity range. In contrast, while the stellar mass function is incomplete for low-mass galaxies, it remains complete for the high-mass regime that defines the LRG population. This completeness at high masses allows us to reliably model the incompleteness at lower masses, making stellar mass a more practical and robust choice for abundance matching LRGs.

To construct the \textsc{Uchuu}-LRG lightcones, we follow the SHAM approach introduced in section 4.1 of \cite{2016MNRAS.460.1173R}, which has been previously applied to the BOSS survey and the DESI Early Data Release  \citep{2023arXiv230606315P}. This method assumes that the most massive galaxies are hosted by the most massive (sub)haloes. The main equation of the SHAM method is

\begin{equation}
    n_{g}(>M^{*}) = n_{h}(>V_{\rm peak}^{\rm scatter}),
\end{equation}
with 

\begin{equation}
    V_{\rm peak}^{\rm scatter} = (1+\mathcal{N}(0, \sigma^{2}))V_{\rm peak},
    \label{scatter_LRG_eq}
\end{equation}
where V$_{\rm peak}$ is the maximum circular velocity over the halo history, $\mathcal{N}$(0, $\sigma^{2}$) is a normal distribution with mean equal to zero and standard deviation $\sigma$, which is the only free parameter of the model. $n_{g}$ is the theoretical (complete) stellar mass function from the LRG population. 

In this work, we found that the optimal value of $\sigma$ depends on the redshift and on the stellar mass. This dependence was found matching the mock two-point correlation function (TPCF) with that from the data split in the redshift ranges $0.4<z<0.6$, $0.6<z<0.8$ and $0.8<z<1.1$ and also doing cuts in stellar mass ($\log{M*}>10.8$, 11.2, 11.3, 11.4 and 11.6). The optimal values of the scatter with the stellar mass and redshift can be seen in Table \ref{scatter_LRG}.

\begin{table}[]
\caption{Value of the scatter found for each cumulative cut in the stellar mass (first column) for each of the three redshift ranges: $0.4<z<0.6$ (second column), $0.6<z<0.8$ (third column) and $0.8<z<1.1$ (fourth column).}
    \centering
    \begin{tabular}{c|ccc}
    \toprule
       $>\log{M^{*}}$  &  $0.4<z<0.6$ &  $0.6<z<0.8$ &  $0.8<z<1.1$\\
       \midrule
       11.6 & 0.62 & 0.58 & 0.30\\
       11.4 & 0.66 & 0.37 & 0.70\\
       11.3 & 0.29 & 0.33 & 0.55\\
       11.2 & 0.14 & 0.1 & 0.16 \\
       11.1 & 0.14 & 0.1 & 0.16\\
       10.8 & 0.14  & 0.1 & 0.16 \\
       \bottomrule
    \end{tabular}
    \label{scatter_LRG}
\end{table}

\subsubsection{Stellar Mass Function and Completeness Modelling}
In the right panel from Figure \ref{LF_SMF} we present the stellar mass function of LRG obtained from the Y1 DESI survey using the \textsc{CIGALE} tool \citep{2019A&A...622A.103B} to estimate individual stellar masses \cite[see][for more details]{2024arXiv240919066S}. We had to use the SMF from Y1 DESI LRG as there are no estimations for the stellar masses of Y3 LRG so far. However, we don't expect a big change in the SMF from Y1 LRG and Y3 LRG.

With this estimations of the stellar masses, we are able to calculate the complete LRG stellar mass function (SMF) at high masses, but we lack information on the shape of the SMF at low masses due to the selection function. To supplement our analysis, we incorporate the SMF measurements obtained from PRIMUS presented by \cite{2013ApJ...767...50M}, which is shown by the square symbols in the right panel from Figure \ref{LF_SMF}. It is worth noting that we do not consider the redshift evolution of the PRIMUS SMF in our analysis, as it has been shown to have a negligible impact on our results and is consistent with the findings of the PRIMUS survey. 

To account for the observed evolution in the shape of the SMF with redshift (as shown in the right panel from Figure \ref{LF_SMF}), we have employed two different complete SMF fits in our SHAM method: one that characterizes a complete galaxy population at 0.4$\leq z \leq$0.6 (low-$z$, represented by the dashed line), and another that characterizes a complete population at 0.6$\leq z \leq$0.85 (high-$z$, represented by the dotted line). 

We apply SHAM to the (sub)halo snapshots from \textsc{Uchuu} boxes at redshifts 0.49, 0.56, 0.63, 0.70, 0.78 and 0.86, 0.94 and 1.03 to cover interval 0.4$\leq z \leq$1.1. As we only have the complete stellar mass function calculated at $0.4<z<0.6$ and $0.6<z<0.85$, and we have halo boxes with redshift beyond $z=0.85$, we apply the complete stellar mass function that corresponds to the range $0.6<z<0.85$ to the two halo boxes with redshift $z>0.85$.

\subsubsection{Incompleteness Implementation: Detailed Procedure}

It is important to remark that one additional step is needed in order to reproduce the observed stellar mass function (points) from Figure \ref{LF_SMF}: we need to randomly remove some galaxies, specially low-mass galaxies, so that we can reproduce the incompleteness that comes from observations. This random downsampling is done for each mass bin in the observed stellar mass function and in small redshift ranges (from 0.4 to 1.1 in steps of 0.01). This probabilistic treatment preserves the clustering properties of the high-mass population while realistically representing the observational limitations. Also, matching the observed stellar mass function we guarantee that the observed number density is also matched.

In the right panel from Figure \ref{HOD} the halo occupancy distribution (HOD) from \textsc{Uchuu}-LRG galaxy box at redshift $z=0.78$ is shown. Again, we can compare the HOD from this work with that presented in \cite{2023arXiv230606315P}. It can be checked that the HOD for central galaxies agrees perfectly well within 1$\sigma$ with that from \cite{2023arXiv230606315P} for the most massive haloes, although there is a big discrepancy for the least massive part of the HOD and for the satellite HOD. Therefore, the full HOD doesn't agree with the the HOD from \cite{2023arXiv230606315P}. As in the case for the BGS, the main difference between the methodology in this work and in \cite{2023arXiv230606315P} is that in the latter work a single scatter in V$_{\rm peak}$ is applied, whereas in this work we apply different scatters for different galaxies depending on the mass. This could explain the difference observed in the HOD.

\subsection{Constructing the \textsc{Uchuu} lightcones}
After applying the SHAM method to populate the simulation catalogues with galaxies, we generate \textsc{Uchuu}-DESI lightcones for each tracer by assembling cubic simulation boxes into spherical shells \cite[see][for a detailed explanation of this method]{2022MNRAS.516.1062S, 2023arXiv230606315P}.  

This approach was applied to both DESI tracers considered in this work to construct their respective full-sky lightcones. The lightcones were built using shells with a redshift width of 0.05, selecting the simulation box with a redshift closest to the mean redshift of each bin.

The shell assignment process means that the same simulation snapshot may be used for multiple adjacent shells when it represents the closest available redshift. For instance, a simulation box at z=0.49 might be assigned to shells covering both $0.45<z<0.50$ and $0.50<z<0.55$. This results in the same galaxies appearing in the two shells, which is a necessary consequence of using finite box sizes to construct continuous lightcones. However, this overlap preserves large-scale structure continuity and does not significantly bias our clustering statistics, as the measurements average over large volumes and many galaxy pairs.

The lightcones were then cut to match the northern and southern areas of the Y1 and Y3 DESI Survey footprints. In this study, we retained all objects within the survey footprint, regardless of completeness of the footprint, for both tracers.

\begin{figure*}
    \centering
    \includegraphics[width=1.0\linewidth]{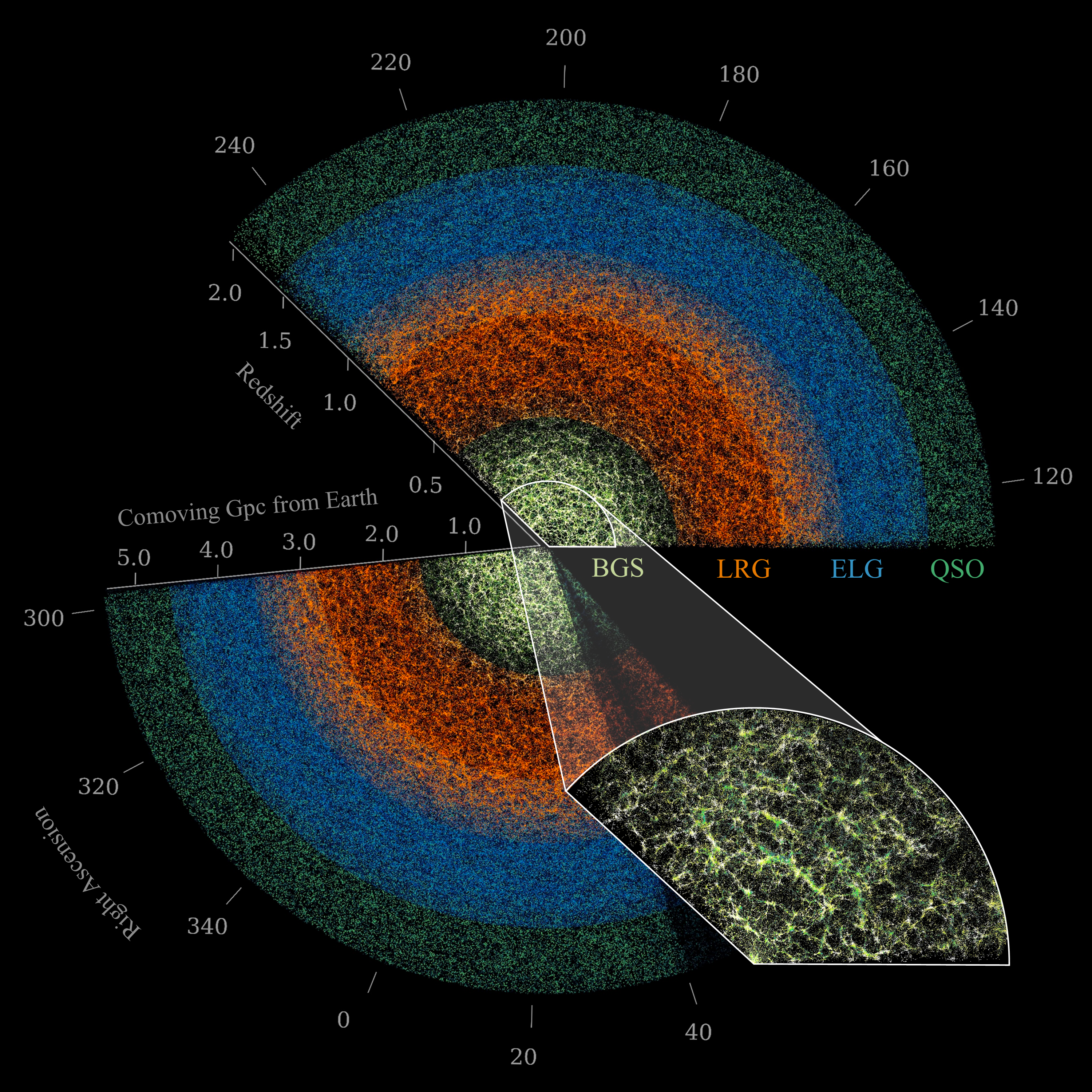}
    \caption{A slice of the universe mapped by Uchuu DR2 drawn from a small wedge of the DESI footprint between $\pm$5 degrees in declination out to $z\approx4$. We render the four major extragalactic samples—bright galaxy survey (BGS) galaxies, luminous red galaxies (LRG), emission– line galaxies (ELG), and QSOs—using yellow, orange, blue, and green points, respectively. Within each target class, the shade of the colour maps to declination (lighter colours correspond to higher declination). The inset shows a subset of the BGS survey extending out to redshift $z = 0.2$, highlighting the large-scale structure traced by galaxies in the densest survey region. For reference, this small wedge of the BGS survey represents less than 0.1$\%$ of the comoving cosmological volume in DR2.}
    \label{claire}
\end{figure*}

Figure \ref{claire} illustrates the incredible scope and detail of the data included in Uchuu-DESI DR2. In this plot a two-dimensional projection of redshift and right ascension for a narrow wedge of the DESI DR2 footprint (±5 degrees in declination) is shown, unveiling the large-scale structure traced by the BGS, LRG, ELG, and QSO targets out to $z \approx 2$.

\section{Redshift-space two-point correlation function}\label{results}
In this section, we compare the two-point correlation function (TPCF) measured for each of the galaxy samples in the DESI DR2 Survey \citep{2024arXiv241112020D} with that predicted by the Planck cosmology using our \textsc{Uchuu} Y3 mock lightcones from LRG and BGS-BRIGHT, as described in the previous section. Additionally, we explore the dependence of the galaxy clustering on luminosity and stellar mass for BGS and LRG galaxies, respectively. 

The LRG data is primary weighted using the Pairwise Inverse Probability (PIP) weights \citep{2017MNRAS.472.1106B}, which reflect whether pairs of galaxies would have been observed in 128 alternate realizations of DESI, in addition to which pairs are observed in the actual survey \citep{2024arXiv240403006L}. The PIP weights are combined with the FKP weights \citep{1994ApJ...426...23F}, $w_{\rm FKP}$, to account for the inhomogeneous sampling density of the data sample, which are defined as

\begin{equation}
    w_{\rm FKP} = \frac{1}{1+P_{0}n(z, n_{\rm tile})}
\end{equation}
where $n(z, n_{\rm tile})$ is the weighted number density, $n_{\rm tile}$ is the number of overlapping tiles, and $P_{0}$ is the power spectrum value at a scale of $k=0.15$ $h$Mpc$^{-1}$ (7000 $h^{-3}$Mpc$^{3}$ for BGS and 10000 $h^{-3}$Mpc$^{3}$ for LRG \citep{2024AJ....168...58D}).


FKP weights are designed to optimize the signal-to-noise ratio by down-weighting high-density regions and up-weighting low-density ones, effectively flattening the redshift distribution $n(z)$. In the case of the BGS-BRIGHT sample (with $M_{\rm r}<-20.0)$, which is flux-limited, $n(z)$ varies significantly with redshift. While FKP weights can still be applied in such cases to mitigate this variation, doing so would alter the effective $n(z)$ of the data relative to that of the mocks, where no FKP weights have been applied. To maintain consistency between data and simulations and avoid introducing artificial differences in clustering—especially at high redshift where $n(z)$ is lower—we do not apply FKP weights when analysing the full, flux-limited sample.

We use the Landy-Szalay \citep{1993ApJ...412...64L} estimator to measure the two-dimensional correlation function, $\xi(s,\mu)$, where $s$ is the pair separation in units of $h^{-1}$Mpc, and $\mu$ is the cosine of the angle between the separation vector and the line of sight.

For large-scale clustering analysis, we use linearly spaced separation bins from 50 to 200 $h^{-1}$Mpc with a bin width of 4 $h^{-1}$Mpc. For small-scale studies, we adopt 49 logarithmically spaced bins between 0.01 and 100 $h^{-1}$Mpc, although our plots primarily focus on the smaller scales. However, the full measurement range will be publicly available. In both cases, we use 30 bins for $\mu$, spanning from -1 to 1.

Additionally, when studying the Baryonic Acoustic Oscillation (BAO) scales we will compare the mock predictions with the measurements from DESI DR1, and when studying the smaller scales we will compare the mock prediction with the measurements from DESI DR2.

We measure the monopole and quadrupole, which are the first non-zero Legendre multipoles of the redshift-space TPCF. To account for the selection function, we generate our random samples that are 16 times larger than the data and use them to estimate the data-random and random-random pair counts for each tracer. We estimate the TPCF using the Python package \textsc{pycorr}\footnote{\url{https:://github.com/cosmodesi/pycorr}} \citep{2020MNRAS.491.3022S, 10.1007/978-981-13-7729-7_1}, which is a wrapper for correlation function estimation wrapping a modified version of \textsc{Corrfunc} \citep{2020MNRAS.491.3022S}. For the \textsc{Uchuu} lightcones, we create a different set of uniform randoms for each tracer, which match the same footprint as the mocks.

\begin{figure*}
    \centering
    \includegraphics[width=0.494\linewidth]{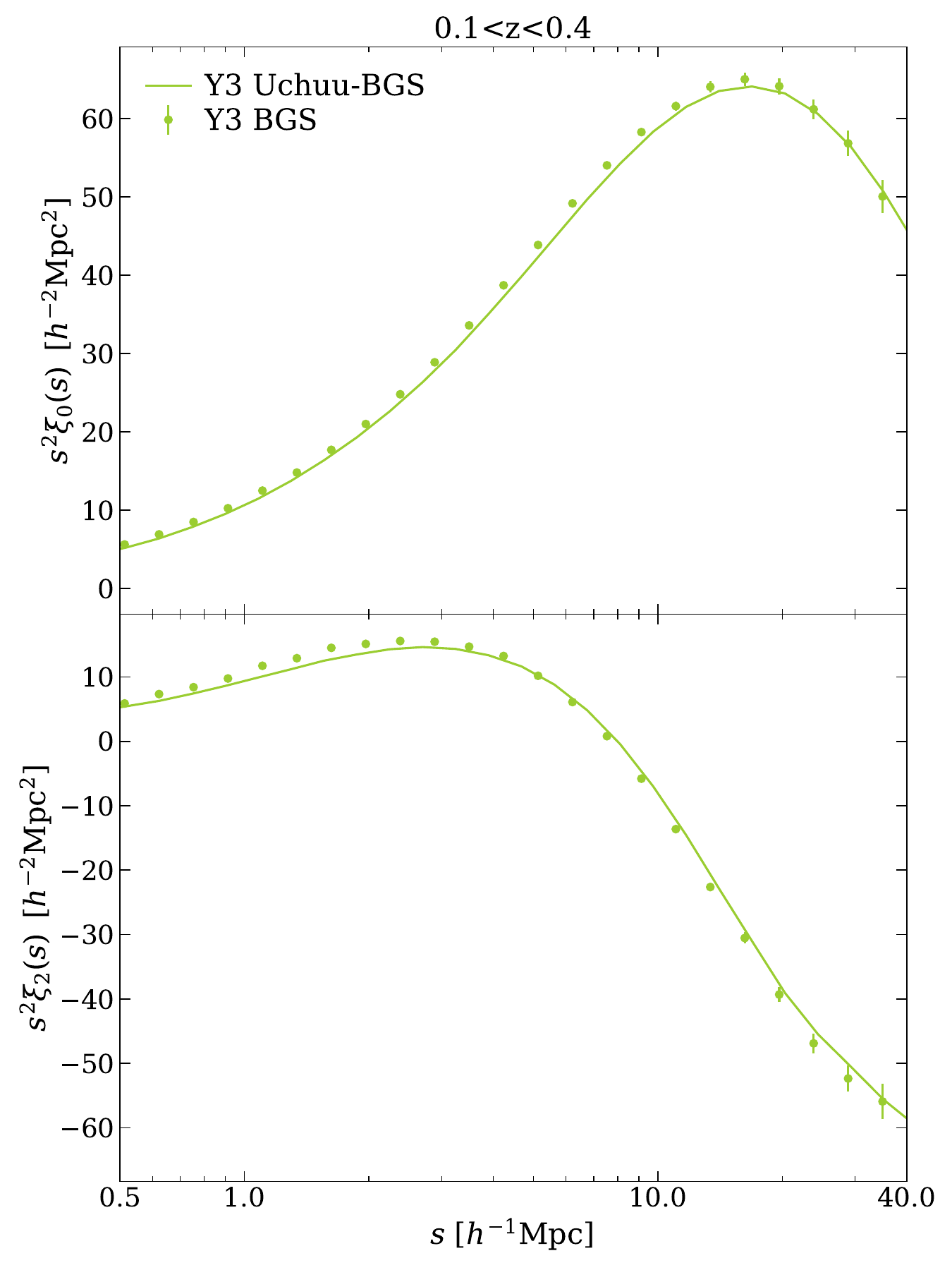}
    \includegraphics[width=0.49\linewidth]{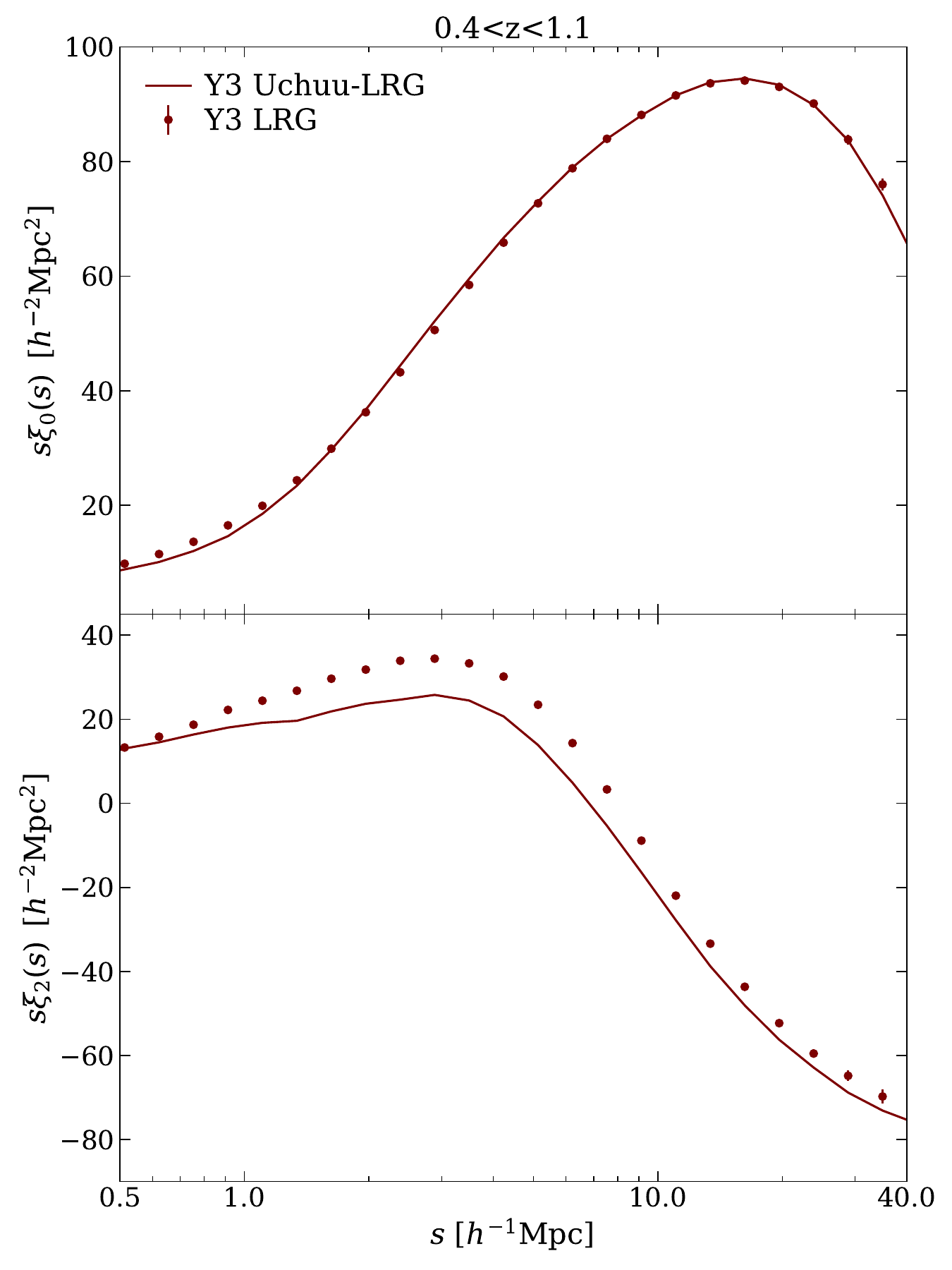}
    \caption{The monopole of the two point correlation function obtained for \textsc{Uchuu}-BGS (\textsc{Uchuu}-LRG) and BGS-BRIGHT (LRG) from Y1 in are shown with solid line and points, respectively, in the left (right) plot. In the bottom panels of both plots the quadrupole can be seen.}
    \label{2pcf_y3}
\end{figure*}

The left panel of Figure \ref{2pcf_y3} shows the monopole (top) and quadrupole (bottom) of the two-point correlation function (TPCF) for Y3 \textsc{Uchuu}-BGS (lines) and Y3 BGS-BRIGHT (points).  

For the monopole, the mean deviation between the mock and the data is approximately 10\% for scales in the ranges $1<s<5$ $h^{-1}$Mpc and $5<s<20$ $h^{-1}$Mpc, decreasing to 5\% for $0.1<s<1$ $h^{-1}$Mpc.  

For the quadrupole, the mean deviation is around 4\% for $1<s<5$ $h^{-1}$Mpc, decreasing to 2\% for $5<s<20$ $h^{-1}$Mpc, but increasing to about 5\% for the smallest scales, $0.1<s<1$ $h^{-1}$Mpc.

In the right panel of Figure~\ref{2pcf_y3}, we show the two-point correlation function (TPCF) for the full sample of luminous red galaxies (LRGs) with redshifts in the range $0.4 < z < 1.1$. The monopole (upper panel) displays excellent agreement between the mocks and the data, with a mean deviation of only 0.2\% across the scales $1 < s < 5~h^{-1}\mathrm{Mpc}$ and $5 < s < 20~h^{-1}\mathrm{Mpc}$. At smaller scales, $0.1 < s < 1~h^{-1}\mathrm{Mpc}$, the deviation increases to around 10\%. 

In the lower panel, which shows the quadrupole, a noticeable discrepancy appears at scales larger than $s = 1~h^{-1}\mathrm{Mpc}$. This mismatch is very likely driven by stellar-mass incompleteness in the sample, which affects in particular the less massive galaxies and has a direct impact on redshift-space clustering. This interpretation is supported by the fact that the agreement in the quadrupole improves significantly for more massive samples, as shown below. 

In addition, the wide redshift bins used in the analysis—especially at high redshift, where the redshift distribution $n(z)$ departs significantly from being constant—can further amplify the effect of stellar-mass incompleteness on the quadrupole. It is worth noting that the agreement in the lower redshift bin, $0.4 < z < 0.6$, is very good, and that the quadrupole measurements for the various volume-limited samples considered in this work for BGS-BRIGHT also show good agreement.

\begin{figure*}
    \centering
    \begin{tabular}{cc}
        \includegraphics[width=\linewidth]{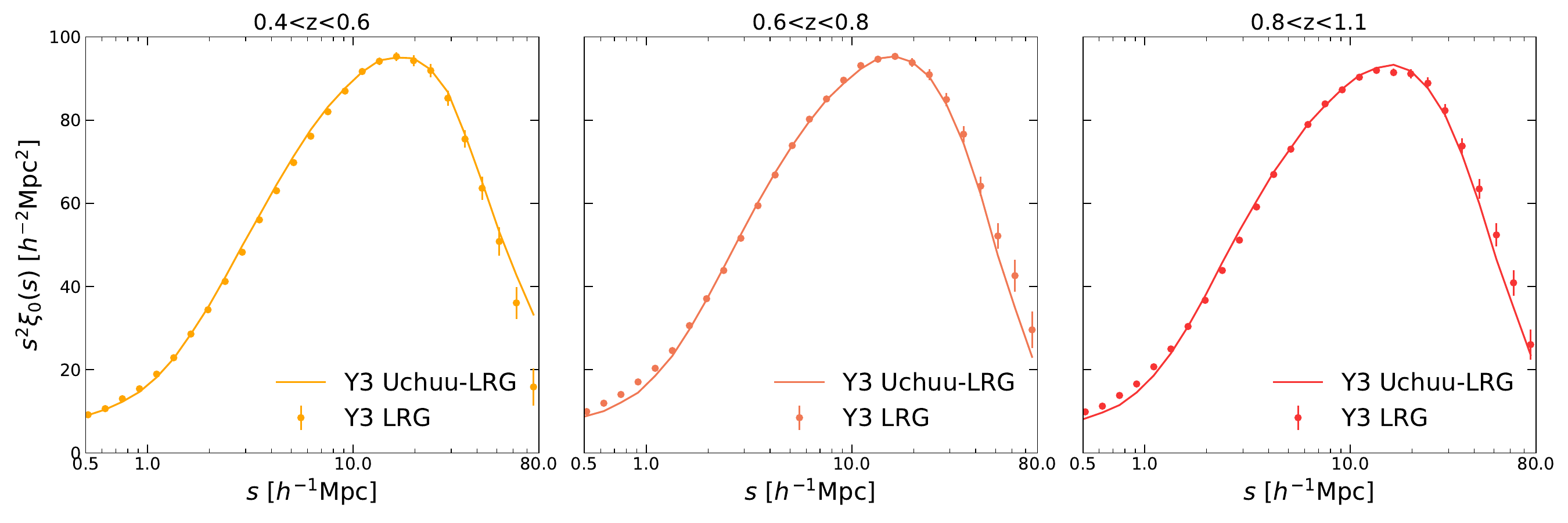}\\
        \includegraphics[width=\linewidth]{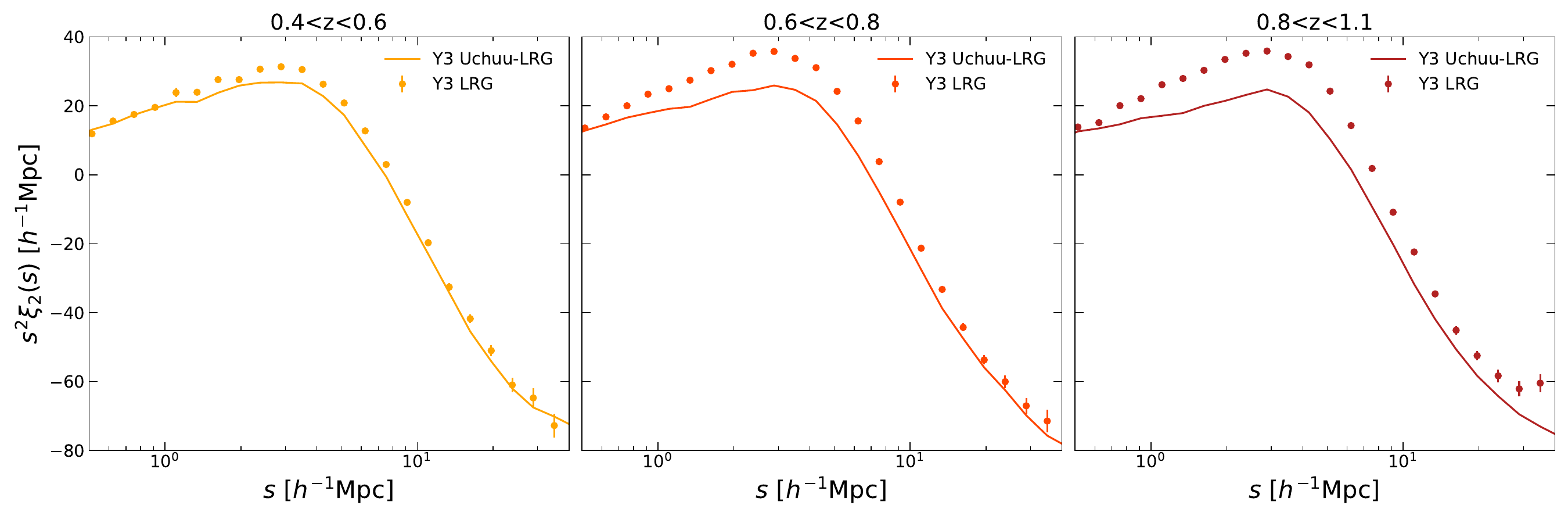}
    \end{tabular}
    \caption{The monopole (quadrupole) of the two point correlation function obtained for \textsc{Uchuu}-LRG (solid lines) and LRG (points) Y1 is shown in the top (bottom) panel for three different redshift ranges ($0.4<z<0.6$, $0.6<z<0.8$, and $0.8<z<1.1$).}
    \label{2pcf_lrg}
\end{figure*}

We can also study the dependence of the TPCF with redshift. This can be seen in Figure \ref{2pcf_lrg}, where the monopole of the TPCF from Y3 \textsc{Uchuu}-LRG (lines) and Y3 LRG (points) for the redshift ranges $0.4<z<0.6$ (first row), $0.6<z<0.8$ (second row) and $0.8<z<1.1$ (third row) is shown. From these figures, one can see that the difference in the monopole between the measurement from the data and the prediction from the mock is below a 2$\%$ for the three redshift ranges for $0.1<s<5$ $h^{-1}$Mpc, below a 1$\%$ for $5<s<20$ $h^{-1}$Mpc and about a 15$\%$ for $0.1<s<1$ $h^{-1}$Mpc. 

In the bottom panels from the same Figure, the quadrupole in each of the three redshift ranges is shown. It can be seen that the discrepancy between the mock and the data seen in Figure \ref{2pcf_y3} comes from the two higher redshift ranges.

\begin{figure*}
    \centering
    \begin{tabular}{cc}     
    \includegraphics[width=0.5\linewidth]{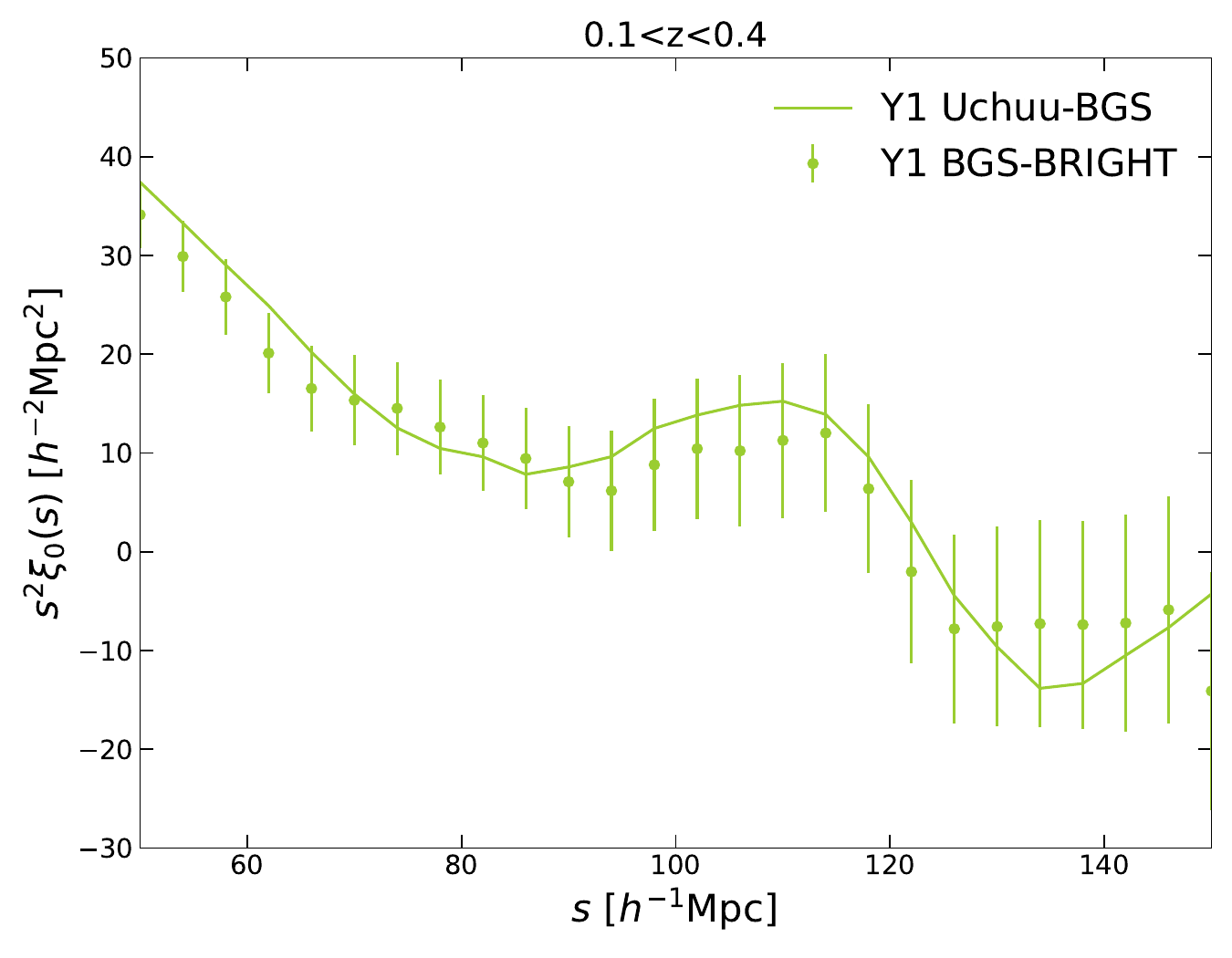}
    \includegraphics[width=0.5\linewidth]{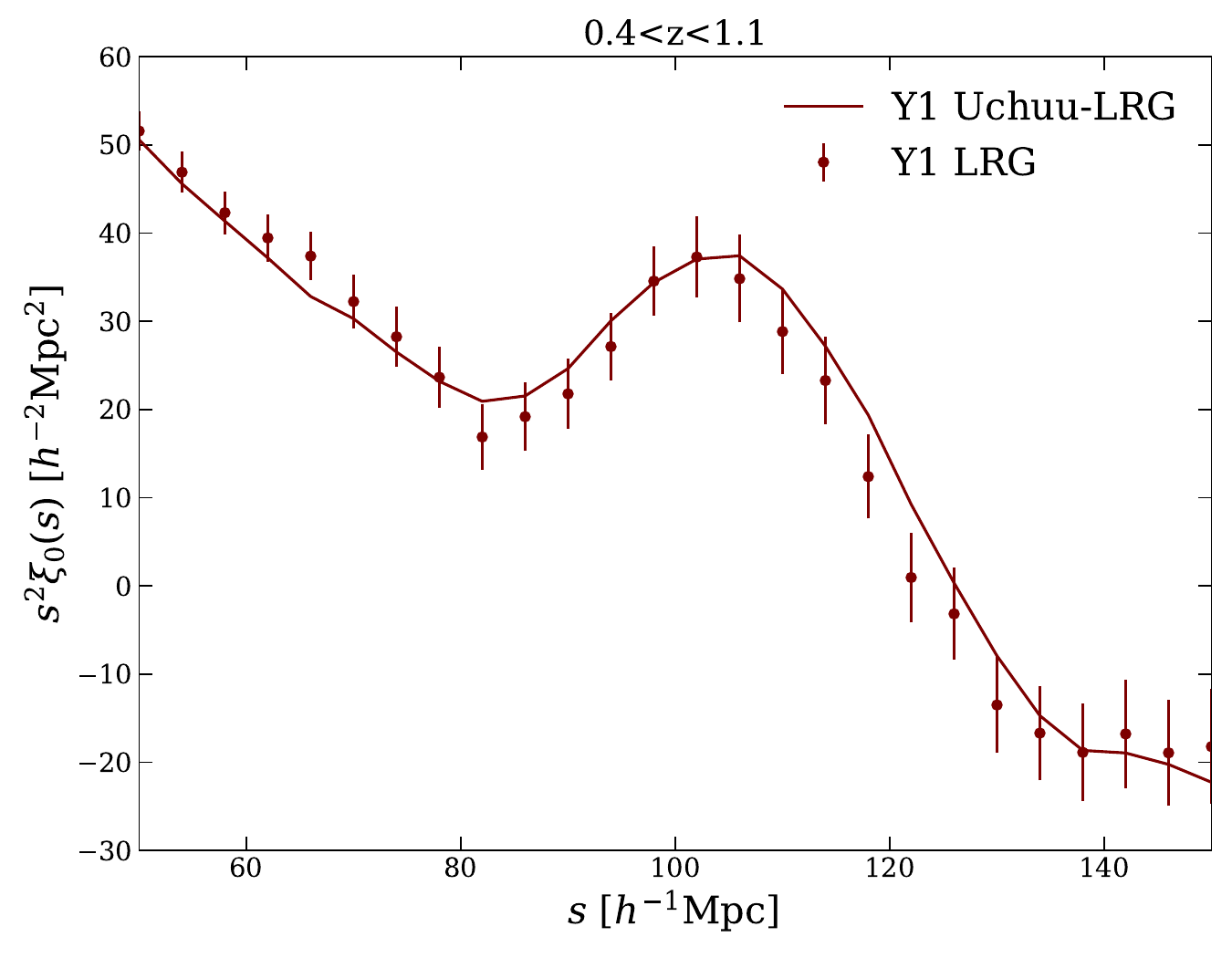}
    \end{tabular}
    \caption{The monopole of the two point correlation function obtained for \textsc{Uchuu}-BGS (solid lines) and BGS-BRIGHT (points) Y1 is shown for the redshift range $0.1<z<0.4$ in the left panel. In the right panel the same can be seen for \textsc{Uchuu}-LRG lightcone and LRG Y1 in the redshift range $0.4<z<1.1$.}
    \label{2pcf_bao}
\end{figure*}

Next, we focused on large scales to study the Baryonic Acoustic Oscillation (BAO) feature in the monopole of the TPCF. A comparison of the TPCF for large scales is shown in the left (right) panel of Figure \ref{2pcf_bao} for the BGS-BRIGHT (LRG) tracers. For both tracers, we observe that the agreement between the data and the mock prediction is quite good, and they are compatible within 1$\sigma$. 

\subsection{Clustering dependence on luminosity and stellar mass}

Now, we can study the dependence of the TPCF with cuts in absolute magnitude (BGS-BRIGHT) and stellar mass (LRG). As already mentioned, matching the dependence of the TPCF from the mocks to the measurements from data is accomplished by introducing a dependence of the scatter (see Eqs. \ref{scatter_BGS_eq} and \ref{scatter_LRG_eq}) with the stellar mass and redshift.

\begin{figure*}
    \centering
    \includegraphics[width=0.49\linewidth]{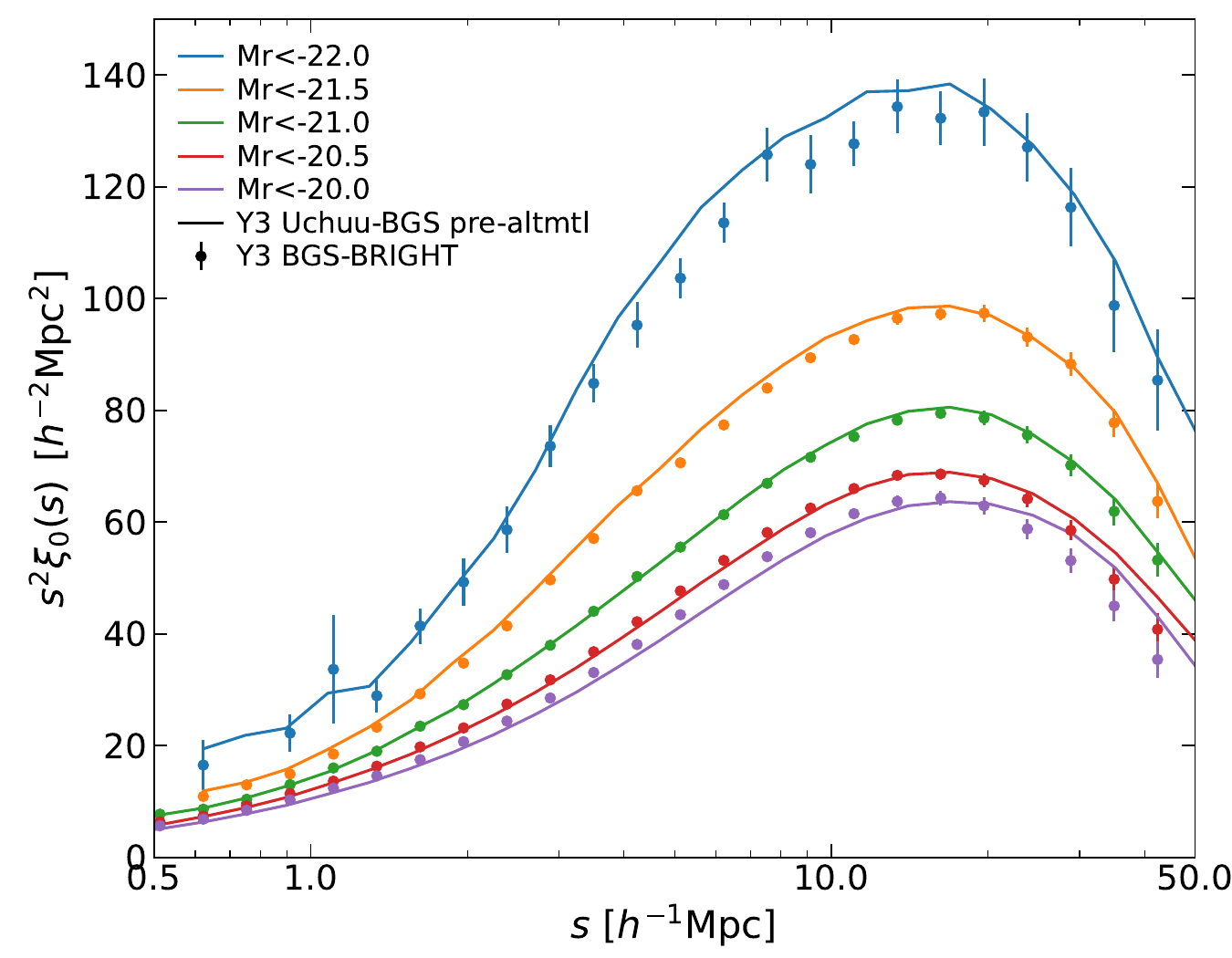}
        \includegraphics[width=0.49\linewidth]{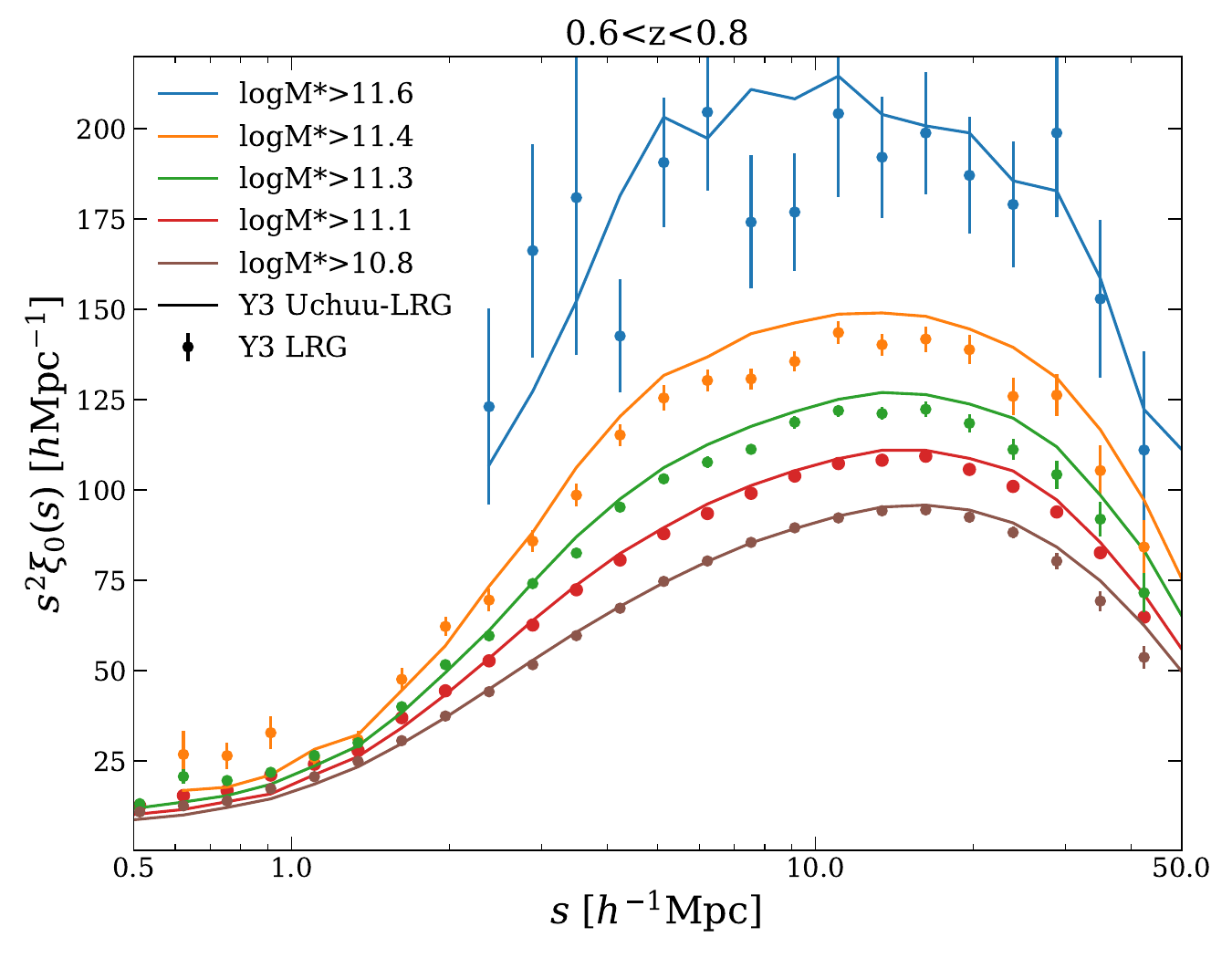}\\
    \includegraphics[width=0.49\linewidth]{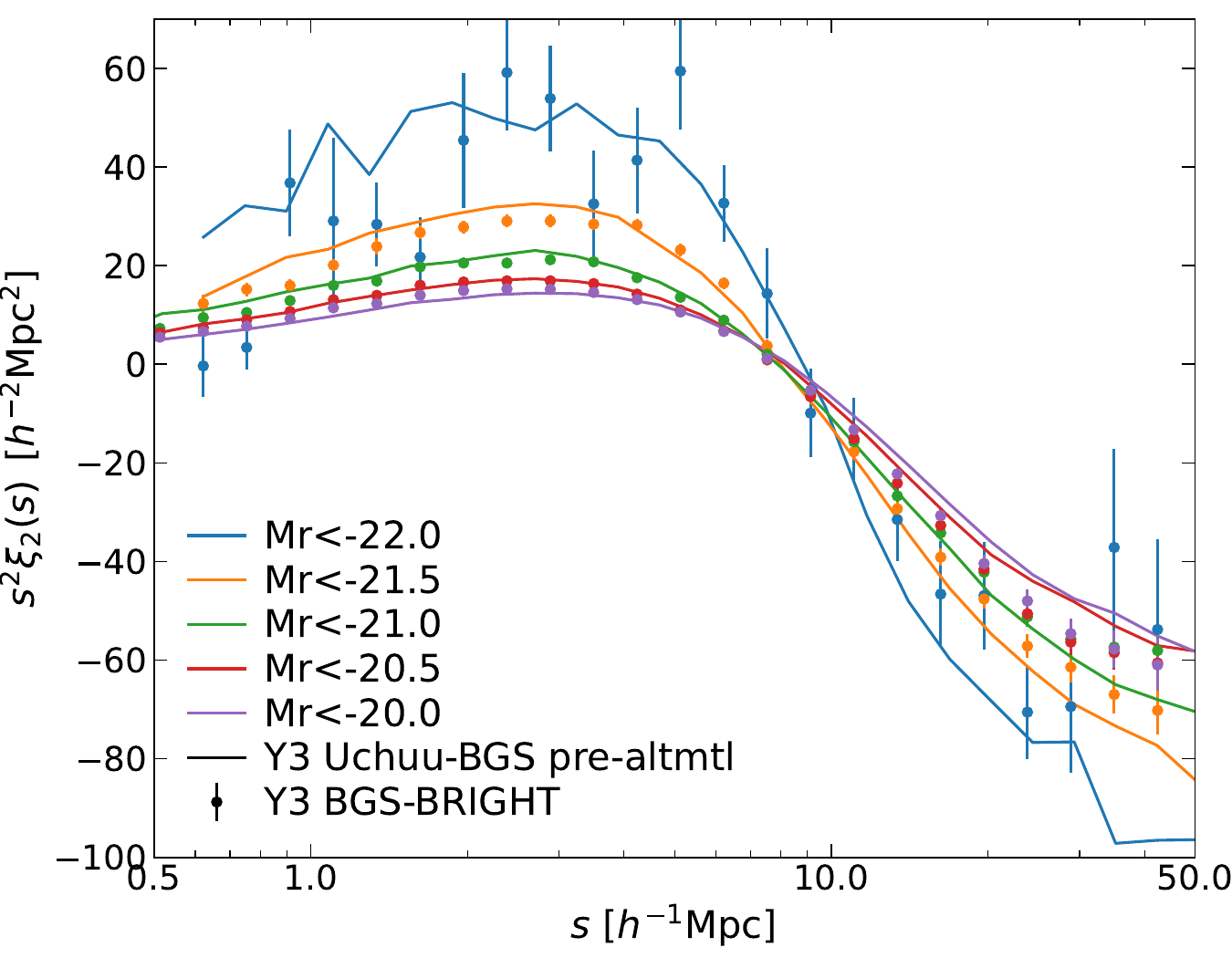}
        \includegraphics[width=0.49\linewidth]{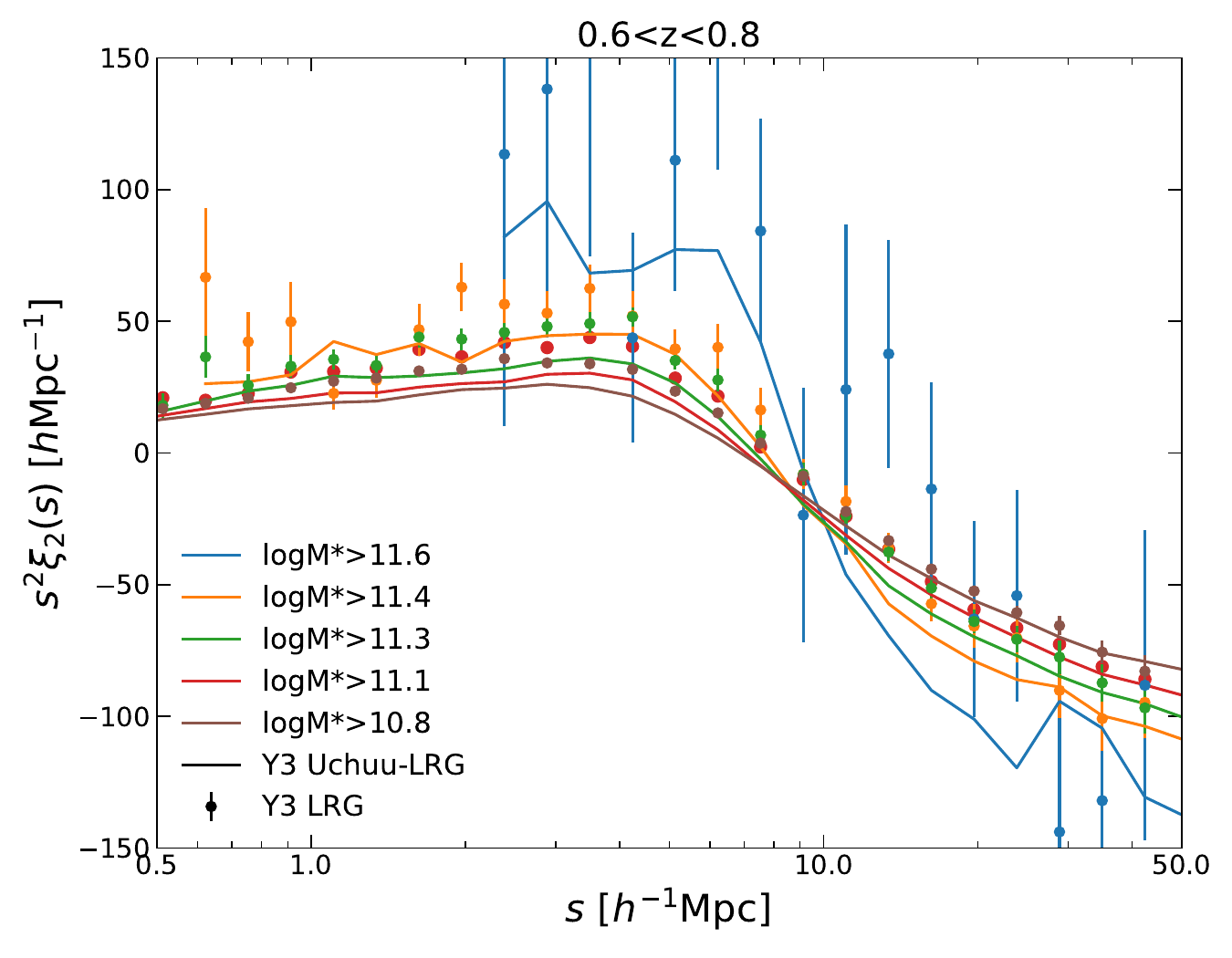}   
    \caption{The monopole of the two point correlation function obtained for the different volume-limited samples detailed in Table \ref{bgs_volsam_tab} from \textsc{Uchuu}-BGS (solid lines) and BGS-BRIGHT (points) is shown in the top-left panel. The same function but for different cuts in stellar mass and for  $0.6<z<0.8$ for the \textsc{Uchuu}-LRG lightcone (solid lines) and LRG Y1 data (points) is shown in the top-right panel. In the bottom panels the quadrupole is shown.}
    \label{2pcf_baryonic}
\end{figure*}

In the top-left panel from Figure \ref{2pcf_baryonic}, one can see the monopole of the TPCF for the different volume-limited samples indicated in Table \ref{bgs_volsam_tab}. The mean deviation between the data and the prediction from the mock for scales between $2<s<30$ $h^{-1}$Mpc is about a 2$\%$ for M$_{\rm r}<-21.5$, below a 1$\%$ for M$_{\rm r}<-21.0$, and -20.5, and about a 2$\%$ for M$_{\rm r}<-20.0$. We can see that there is an increasing deviation between the mock and the data for the faintest volume-limited samples, and this can't be fixed decreasing the scatter value (the monopole of the TPCF increases when decreasing the scatter, as demonstrated in \cite{2024MNRAS.528.7236D}), because the value of the scatter for the faintest galaxies considered in this work is quite close to zero. Therefore, we suspect that the $k+E$ corrections applied to both the data and the mock have to be improved. 

In the bottom-left panel from the same Figure, one can see the quadrupole of the TPCF for the same volume-limited samples. Similar tendencies as in the monopole are observed -- the agreement between the data and hte mock is better for the brightest volume-limited samples.

Next, we studied the dependence of the monopole of the TPCF for different cuts in stellar mass for the LRG tracer. We show the results for the middle redshift range only, $0.6<z<0.8$, but the reader can check using the public catalogues that similar agreement between the data and the mocks is obtained for the other two redshift ranges. The results from the 
\textsc{Uchuu}-LRG lightcone and DESI LRG Y1\footnote{Note that we do the comparison with DESI LRG Y1 because the estimations for the stellar mass using the \textsc{CIGALE} tool have been done only for Y1 so far.} can be seen in the top-right panel from Figure \ref{2pcf_baryonic}. We have calculated that the average deviation of the mocks from the data in the range $5<s<20$ $h^{-1}$Mpc is below a 5$\%$ for all the stellar mass cuts, and between a 10 and a 15$\%$ for scales in the range $0.1<s<1$ $h^{-1}$Mpc. 

Finally, the dependence of the quadrupole with different cuts in the stellar mass for the LRGs can be seen in the bottom-right panel from Figure \ref{2pcf_baryonic}. As expected from Figures \ref{2pcf_y3} and \ref{2pcf_lrg}, the agreement between the data and the mock is not good at small scales, specially for the subsamples that include almost all galaxies (or, in other words, subsamples with a low-massive stellar mass cut).

\subsection{Redshift space distortions}

\begin{figure*}
    \centering
    \includegraphics[width=0.49\linewidth]{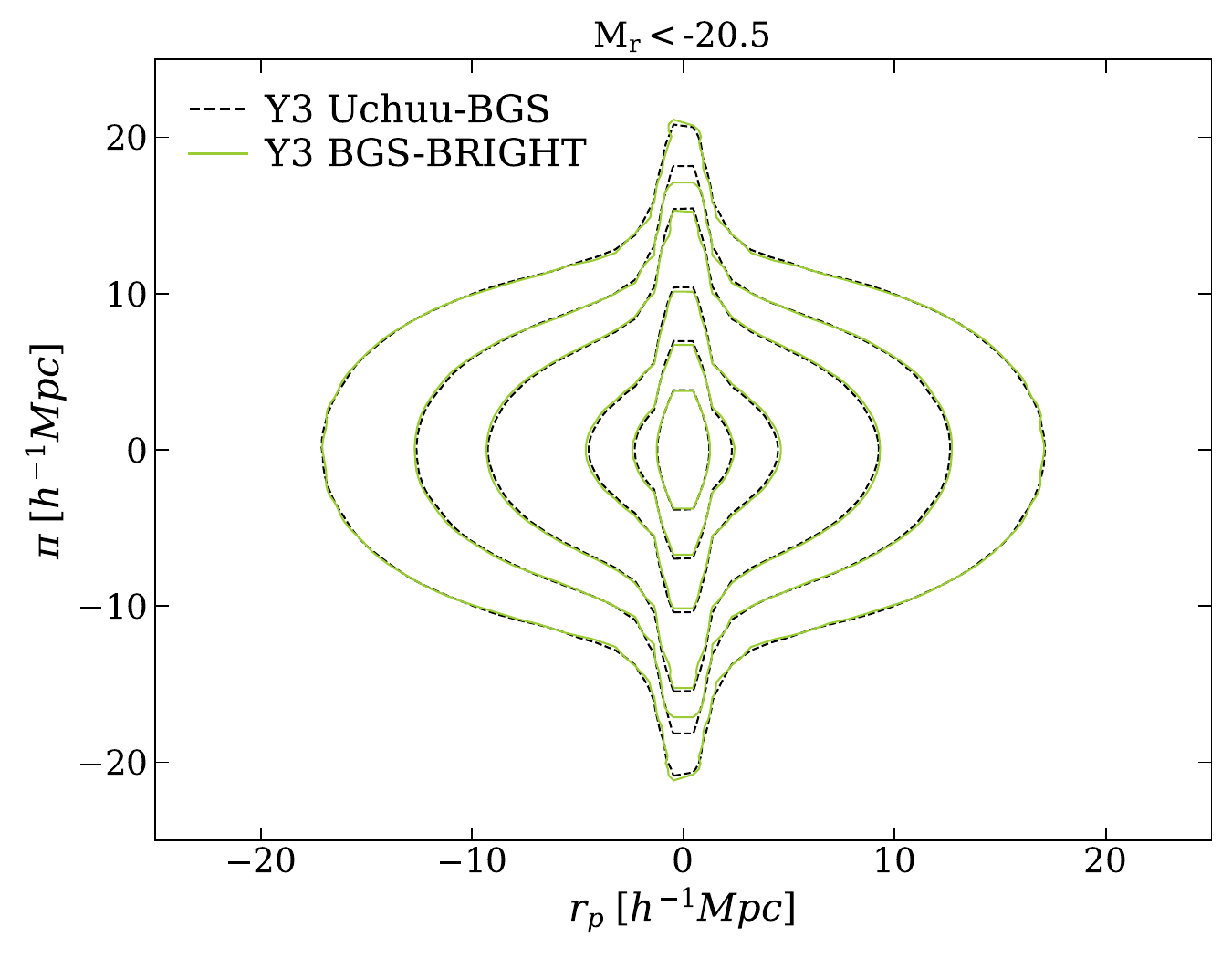}
    \includegraphics[width=0.49\linewidth]{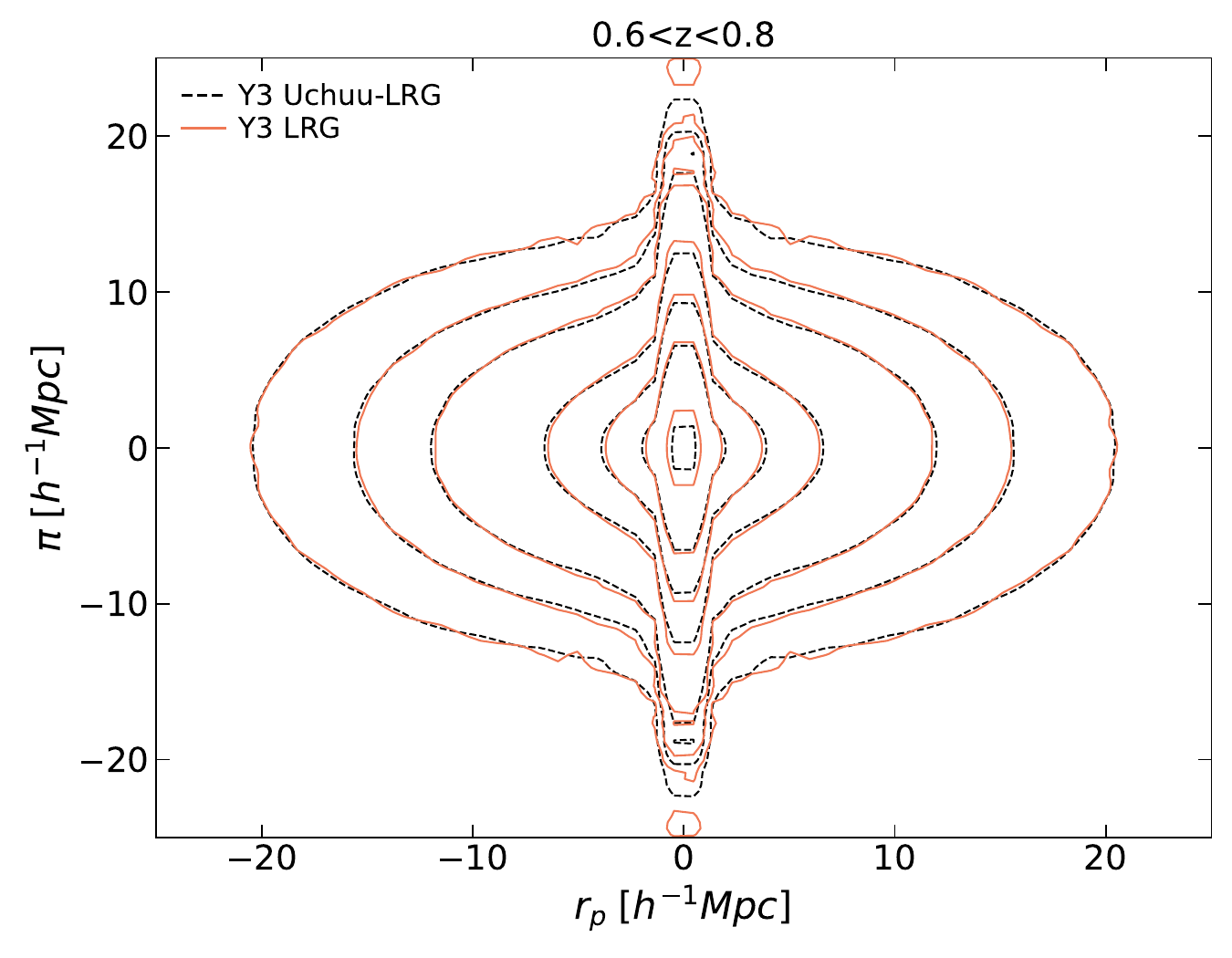}\\
    \includegraphics[width=0.49\linewidth]{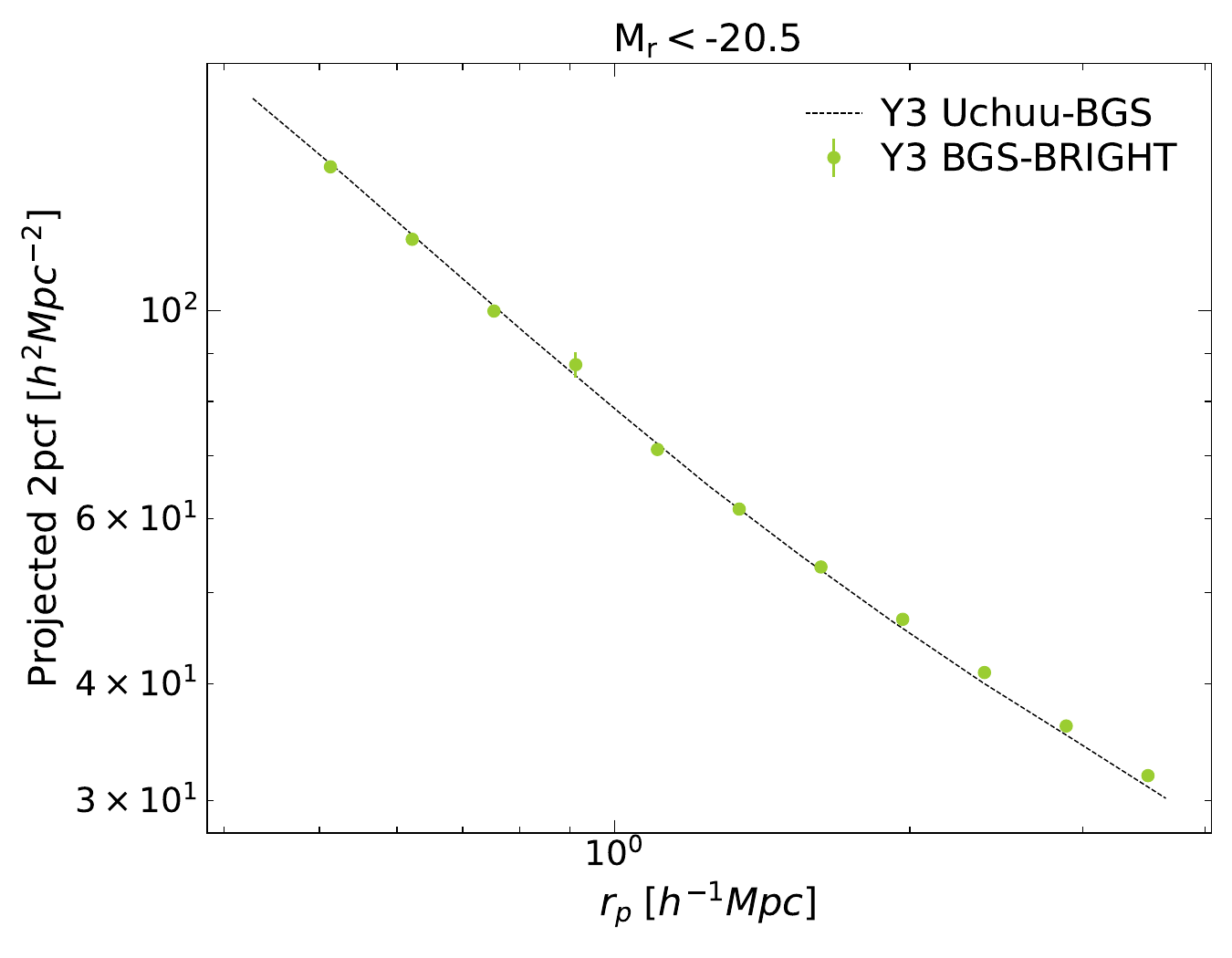}
    \includegraphics[width=0.49\linewidth]{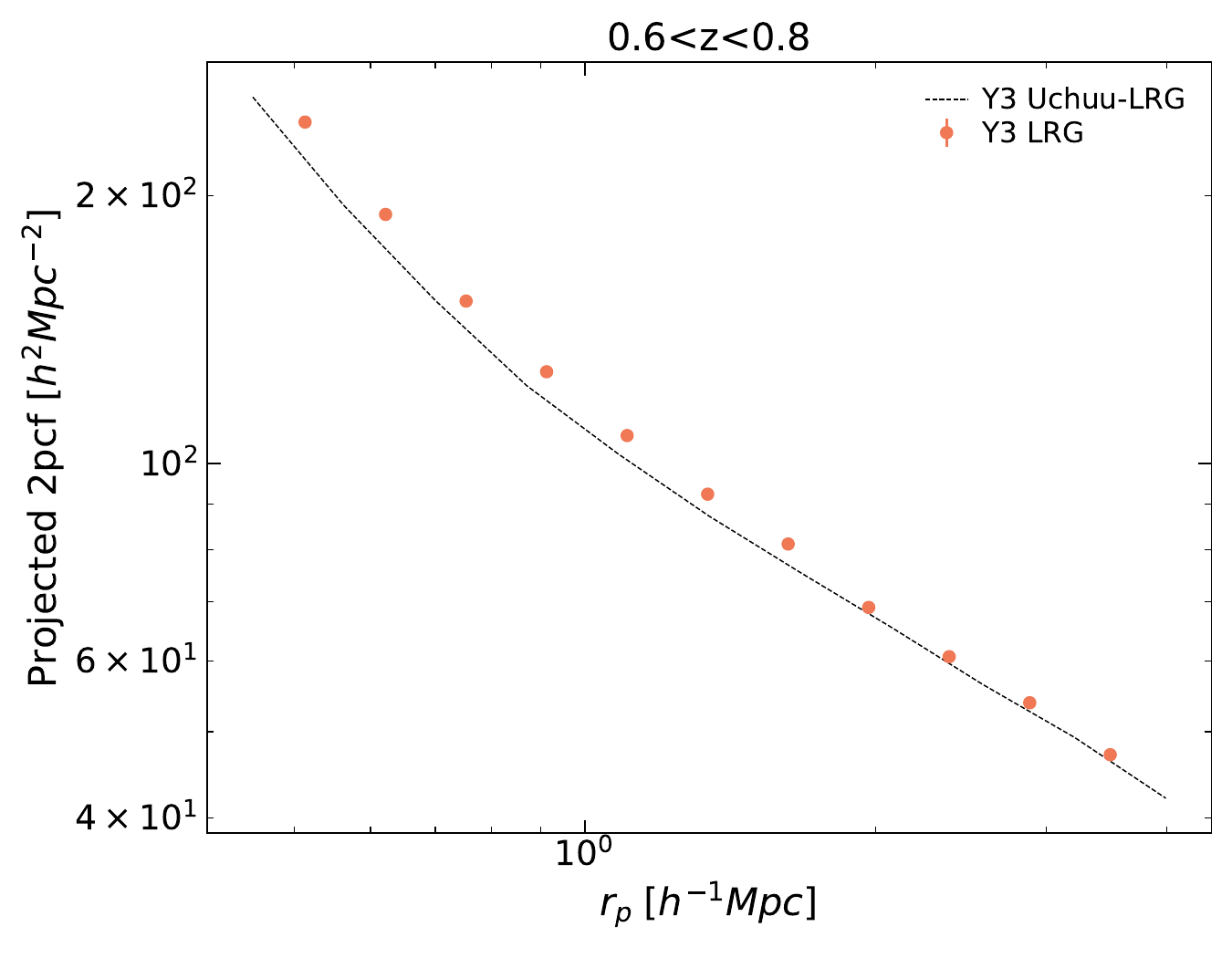}
    \caption{The TPC from \textsc{Uchuu}-LRG (solid line) and Y3 LRG (dashed line) plotted as a function of pair separation perpendicular ($r_{\rm p}$) and parallel ($\pi$) to the line of sight can be seen in the top panel. To illustrate deviations from circular symmetry, the first quadrant is replicated with reflections in both axes. The contours levels shown are [0.5, 0.8, 2, 4, 8, 20]. In the bottom panels we show the same measurements of the projected TPCF integrating $\pi$ out.}
    \label{contours_lrg}
\end{figure*}

Top panels from Figure \ref{contours_lrg} represent the TPCF from \textsc{Uchuu}-BGS and \textsc{Uchuu}-LRG lightcone and BGS-BRIGHT and LRG Y3 plotted as a function of pair separation perpendicular ($r_{\rm p}$) and parallel ($\pi$) to the line of sight. From this Figure the presence of redshift space distortions (RSD) is clear by the elongation along the $\pi$ axis at scales within 5 $h^{-1}$Mpc. The remarkable agreement between our mock and the data is noteworthy.

To facilitate a more precise comparison between the data and the mock on small scales, we have integrated out $\pi$ and projected the measurements onto the $r_{\rm p}$ axis. This is illustrated in the bottom panels of Figure \ref{contours_lrg}. For the case of the BGS (LRG), the discrepancy between the mock and the data is approximately 1 (0.7) $\%$ at $r_{\rm p} = 1.5 h^{-1}$ Mpc and decreases (increases) to about 0.8 (2) $\%$ at $r_{\rm p} = 0.5 h^{-1}$ Mpc. The larger deviation at the smallest scales for the case of the LRGs may be attributed to the absence of a fibre collision correction in our analysis. Additionally, since LRGs are more distant, the angular scales corresponding to the fibre collisions would map to larger physical scales. This may contribute to the observed discrepancies at the smallest scales, as the fibre collision effects are more pronounced for the more distant objects.

\section{Redshift-space power spectrum}\label{power_spectrum}
In this section, we also compare the power spectrum monopole $P_0(k)$ and quadrupole $P_2(k)$ for the galaxy samples considered in the previous section, measured over the wavenumber range $0 < k < 0.2 \,h\,\mathrm{Mpc}^{-1}$ with a bin width $\Delta k = 0.02 \, h\,\mathrm{Mpc}^{-1}$.
The power spectrum estimator is based on the fast Fourier transform (FFT) algorithm, as developed by \citet{2017JCAP...07..002H} and implemented in the Python package \textsc{PyPower}\footnote{\url{https://github.com/cosmodesi/pypower}}.
Incompleteness, redshift failure and imaging systematic weights are applied to the survey data, and FKP weights are calculated and applied to both the survey data and the \textsc{Uchuu} lightcones. 
For each catalogue, the particles are assigned to a mesh grid using the piecewise cubic spline (PCS) scheme, with a minimum grid number in each dimension chosen such that the Nyquist wavenumber is above $0.4 \, h\,\mathrm{Mpc}^{-1}$ to avoid aliasing effects.
Finally, the power spectrum normalisation factor is calculated as a sum over both the galaxy and random mesh grids of grid cell size $10 \, h^{-1}\,\mathrm{Mpc}$, such that it approximates the window function amplitude and thus the power spectrum amplitude can be approximately compared across different survey geometries \citep{2024arXiv241112020D}.

\begin{figure*}
    \centering
    \includegraphics[width=0.49\linewidth]{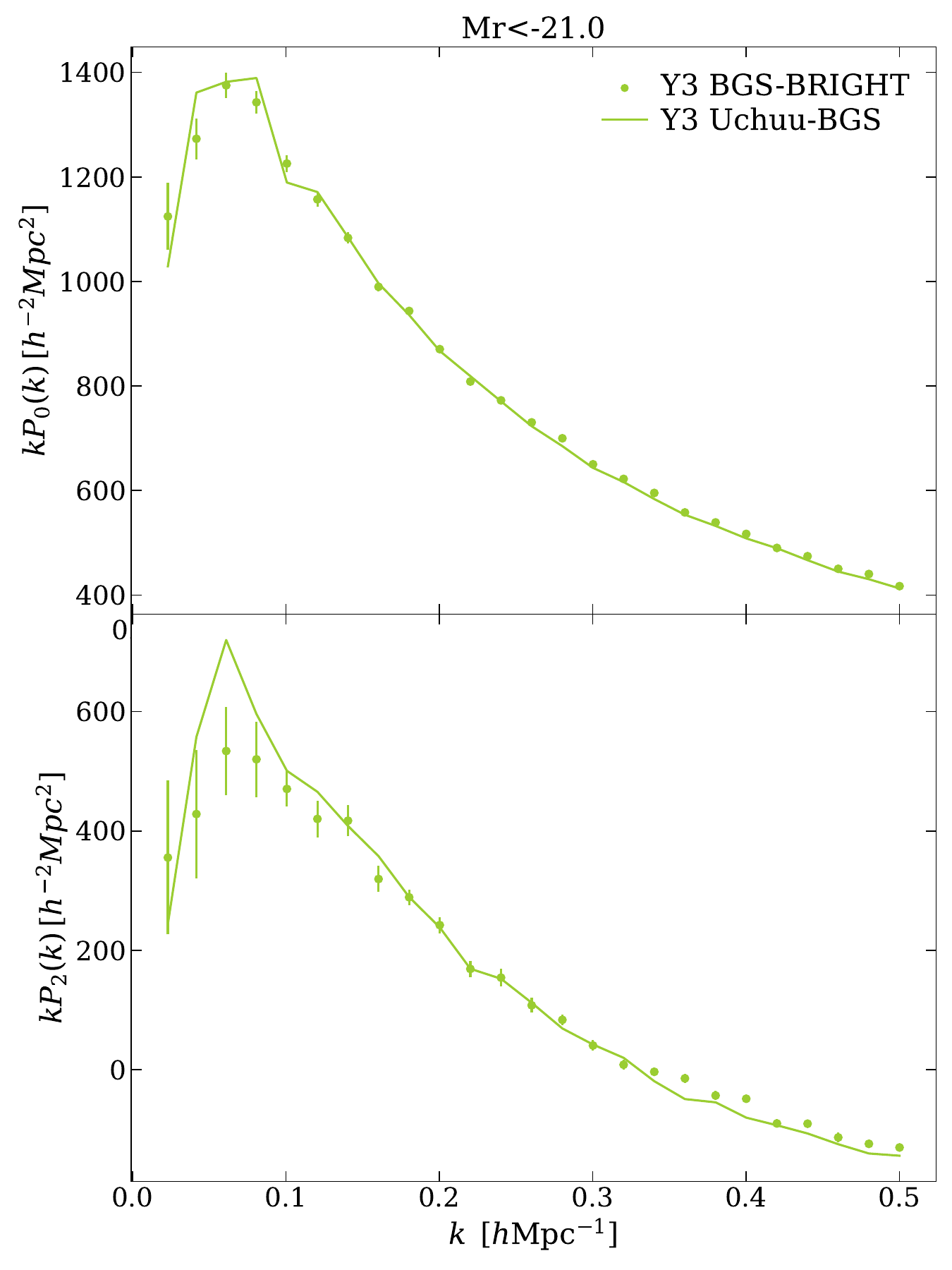}
    \includegraphics[width=0.49\linewidth]{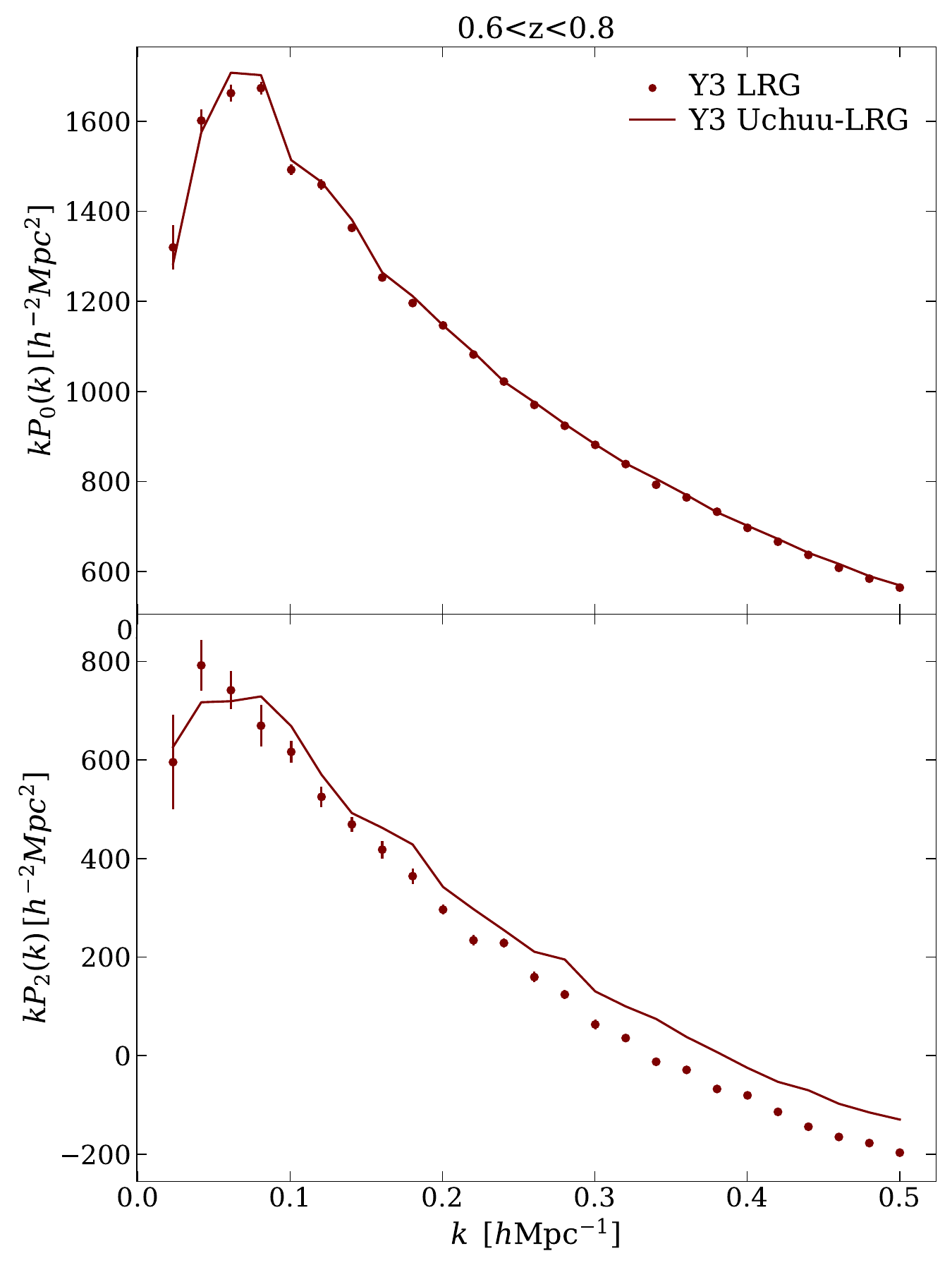}
    \caption{Power spectrum monopole (top panels) and quadrupole measurements (bottom panels) of the BGS-BRIGHT volume-limited sample with M$_{r}<-21.0$ (left) and LRG with $0.6<z<0.8$ (right) tracers. The solid lines and the points show the measurements from \textsc{Uchuu} and the Y3 survey data, respectively. The error bars represent the standard deviation calculated from the 25 AbacusSummit mocks.}
    \label{Pk}
\end{figure*}

In Figure \ref{Pk}, we show the power spectrum monopole and quadrupole measured for BGS volume-limited sample with $M_r < -21.0$ and for LRG in the redshift bin $0.6 < z < 0.8$. We observe excellent agreement between the Y3 survey data and the Uchuu lightcones, especially for LRG. However, one caveat is that the slightly different window functions (of which the various weights are an input) complicate the interpretation as they change the shape of the observed power spectrum \citep{1994ApJ...426...23F}; in addition, the power spectrum normalisation discussed above only approximately accounts for the amplitude modulation induced by the window function.
By contrast, the Landy--Szalay type of TPCF estimators eliminate the effect of the window function.


\section{Large-scale bias}\label{hod_bias}
\begin{figure*}
    \centering
    \includegraphics[width=0.49\linewidth]{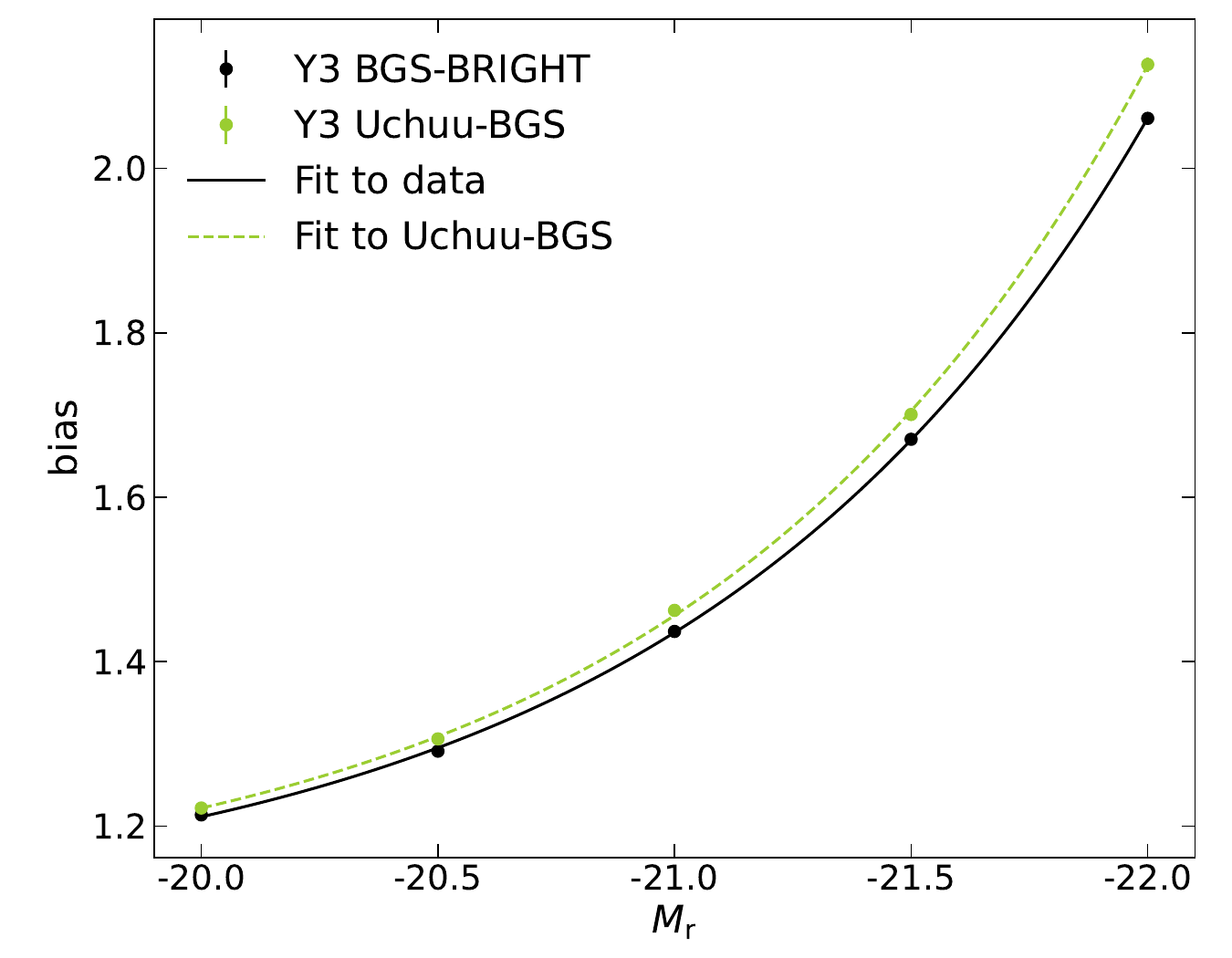}
    \includegraphics[width=0.49\linewidth]{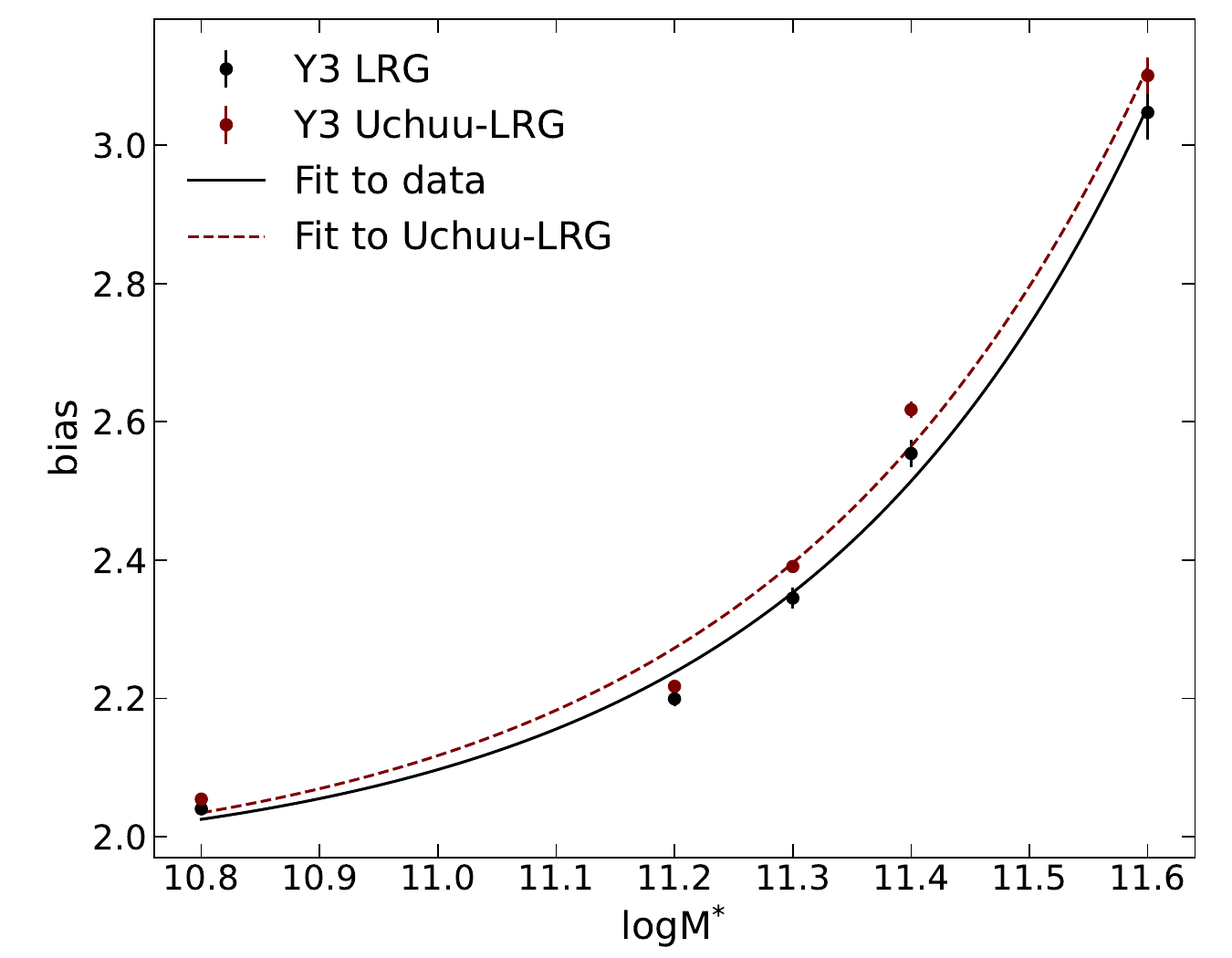}
    \caption{Bias of the BGS (LRG) magnitude threshold volume limited-samples (samples with stellar mass cut), at the median redshift of each sample is shown in the left (right) panel for the data (points) and \textsc{Uchuu} mocks (solid lines).}
    \label{bias_plot}
\end{figure*}

The large-scale bias, $b$, for both DESI tracers was measured from the DESI DR2 Survey and compared to their prediction obtained from the \textsc{Uchuu} lightcones in the Planck cosmology. The results are presented in Figure \ref{bias_plot}. We
measure the linear bias by fitting

\begin{equation}
    \xi_{0}(s) = b^{2}\left(1+\frac{2}{3}\beta+\frac{1}{5}\beta^{2} \right)\xi_{\rm lin}(s)
    \label{xilin}
\end{equation}
to the correlation function monopole measurements, $\xi_{0}(s)$, over
a given range of separations. $\xi_{\rm lin}(s)$ is from the linear power spectrum at the redshift of our galaxy sample, and $\beta$ = $\Omega_{\rm m}^{0.6}/b$ \cite[see][]{1987MNRAS.227....1K, 1998ASSL..231..185H}.

The linear bias for both BGS and LRG samples is measured over the separation range $10 < s < 40$ $h^{-1}$Mpc at the median redshift of each volume-limited sample. For the LRGs, we consider only galaxies in the redshift range $0.6 < z < 0.8$. As expected, the brightest and most massive galaxies exhibit the highest bias, as they reside in the most massive halos, which are more strongly clustered.  

Comparing the bias from Uchuu-BGS to that from BGS-BRIGHT, we find that they are consistent within $1\sigma$ for the faintest volume-limited samples. For the brightest samples with $M_{\rm r} < -21$, the biases show a larger discrepancy. The uncertainties shown for the bias are derived from the fitting procedure and do not include cosmic variance, which can be significant given the finite volume probed by the survey and the simulation. Including cosmic variance would increase the total error budget and could reduce the apparent discrepancy. Despite this, the agreement between mock and data is very good for the 2PCF monopole and quadrupole, the projected 2PCF, and the power spectrum, suggesting that the small difference in bias is not driven by limitations in the SHAM modelling or redshift-space clustering.


A similar trend is observed for the LRGs. In both cases, the bias measured from the mock is higher than that from the data. This is consistent with Figure \ref{2pcf_baryonic}, which shows that the monopole of the TPCF at scales $10 < s < 40$ $h^{-1}$Mpc is larger in the mock than in the data for the brightest and most massive galaxies. Furthermore, at larger scales, the mock predictions systematically exceed the measurements from the data.

The final step we took to study the dependence of the bias with the absolute magnitude and stellar mass is finding a functional dependence between these variables. To do this, we used the same function as that proposed in   \cite{2011ApJ...736...59Z} for the BGS:

\begin{equation}
    b(<M_{\rm r}) = B_{0}+B_{1}\times10^{B_{2}(M_{\rm r}^{*}-M_{\rm r})/2.5}
\end{equation}
where $M_{\rm r}=-20.44$. 

For the LRGs, we use the same equation but change M$_{\rm r}$ by $\log{(M^{*})}$ and fix the value of M$_{\rm r}^{*}$ to 12.3, which is the typical mass for a LRG.

\begin{table}[]
    \centering
    \caption{The table presents the best-fit values of the parameters $B_0$, $B_1$, and $B_2$ in the functional equation of the galaxy bias as a function of luminosity and stellar mass. The first four rows correspond to the Y3 BGS-BRIGHT and Y3 LRG surveys, along with their respective Uchuu-based mock counterparts. The results for both real and mock data are consistent within $1\sigma$, as shown in Figure \ref{bias_plot}. For LRGs, the bias is modelled as a function of $\log{(M^{*})}$, with the characteristic mass fixed at $M_{r}^{*} = 12.3$. The last two rows contain the corresponding values for the DESI One-Percent Survey from \cite{2023arXiv230606315P}. Despite the larger uncertainties in DESI 1$\%$ due to its limited survey volume, the obtained parameters are compatible with the results found in this work.}
    
    \begin{tabular}{c|ccc}
        \toprule
        Survey  & $B_{0}$ & $B_{1}$  & $B_{2}$ \\
        \midrule
        Y3 BGS-BRIGHT & 1.087 $\pm$ 0.008& 0.196 $\pm$ 0.008& 1.12 $\pm$ 0.02\\
        Y3 Uchuu-BGS & 1.10 $\pm$ 0.013& 0.200 $\pm$ 0.012& 1.14 $\pm$ 0.03\\
        Y3 LRG & 1.95 $\pm$ 0.07& 0.40 $\pm$ 0.08& -3.65 $\pm$ 0.55\\
        Y3 Uchuu-LRG  & 1.94 $\pm$ 0.10 & 0.46 $\pm$ 0.12& -3.42 $\pm$ 0.69\\
        \midrule 
        DESI-BGS 1$\%$ & 0.59 $\pm$ 0.43 & 0.62 $\pm$ 0.45 & 0.43 $\pm$ 0.29 \\
        DESI-LRG 1$\%$ & 1.90 $\pm$ 0.018 &  0.26 $\pm$ 0.02 &  -5.69 $\pm$ 0.69  \\
        \bottomrule
    \end{tabular}%
    
    \label{bias_fit}
\end{table}

The values we obtain for $B_{0}$, $B_{1}$ and $B_{2}$ can be seen in the Table \ref{bias_fit}. The results from these two fits can be seen in the left panel from Figure \ref{bias_plot} with continuous lines for Y3 BGS-BRIGHT and Y3 LRG, and dashed lines for the mock.  For both tracers, the results from the mock and the data are compatible within 1$\sigma$. We have also added in the last two rows of the Table the results obtained for the DESI One-Percent Survey in \cite{2023arXiv230606315P}. We can see that the results obtained in that work are compatible with the results obtained in this work, although the uncertainties obtained for the One-Percent survey were much larger because of the tiny volume of that survey.

To conclude this section, we have computed the linear Two-Point Correlation Function (TPCF) using the bias measurements derived earlier for each BGS volume-limited sample and LRG sample with different stellar mass cuts. The results are shown in Figure \ref{linear_tpcf} from Appendix \ref{linear_appendix}, where the left panel displays the linear TPCF at redshift $z=0$ for the data (represented by points) and mocks (depicted by lines) for BGS-BRIGHT targets, while the right panel shows the corresponding data for LRGs. Additionally, the linear TPCF calculated from Uchuu's matter power spectrum is shown as a dark dashed line. The bottom panels of both plots present the ratio between the data and mock linear TPCF. The shaded gray area represents the 5$\%$ limits. From these panels, it is evident that the agreement between the data and mock linear TPCF with that from Uchuu is excellent at scales between $10 < s < 20$ $h^{-1}$Mpc (the ratio is below a 5$\%$), and remains strong even at larger scales. For the calculation of the large-scale bias, we assumed that the TPCF was linear within the range $10 < s < 20$ $h^{-1}$Mpc, and this result confirms the validity of that assumption.

\section{Summary}\label{summary}
This study details the methodology used to generate simulated lightcones for the LGR and BGS-BRIGHT DESI DR2 tracers. These lightcones are built within the flat-$\Lambda$CDM Planck cosmology framework using \textsc{Uchuu}, a 2.1 trillion-particle $N$-body simulation tailored for the DESI survey. \textsc{Uchuu} provides high-resolution modelling of dark matter haloes and subhaloes across a vast volume, capturing structures from massive galaxy clusters to dwarf galaxies.

To populate the \textsc{Uchuu} haloes with DESI galaxies, we apply the Subhalo Abundance Matching (SHAM) technique, using peak maximum circular velocity as a proxy for (sub)halo mass. The construction of BGS and LRG lightcones follows the standard SHAM approach, incorporating the redshift evolution of tracers and the dependence of clustering on key properties such as luminosity and stellar mass.

We compare the predicted clustering signals from \textsc{Uchuu} with measurements from DESI DR2 for each galaxy sample. Additionally, we analyse the halo occupation distribution and large-scale bias factors for the DESI targets.

Our main results are summarized as follows:

\begin{itemize}
    \item We measured the redshift-space two-point correlation function monopole and quadrupole over the scales from 0.01 $h^{-1}$Mpc to 200 $h^{-1}$Mpc. Overall, we find consistency between the DESI DR2 (and Y1) Survey measurements and the theoretical predictions based on Planck cosmology using the \textsc{Uchuu} lightcones. The agreement between the data and the prediction from the mock for both tracers is excellent for the monopole, although we find some discrepancy in the quadrupole for the LRGs, which may be caused to the fact that the SHAM method does not accurately model galaxy velocities. If we focus on BAO scales, we find that the result predicted by the mocks and those measured in DESI DR1 are statistically compatible within 1$\sigma$ for the full sample for both tracers.
    \item We have also shown that we can reproduce quite accurately the monopole of the TPCF for different volume-limited samples for the BGS and different stellar mass cuts for the LRGs. For the BGS, we've been able to reach the limit in absolute magnitude of M$_{\rm r}<-20.0$. We've found that the disagreement between the monopole from the data and that predicted by \textsc{Uchuu}-BGS mocks becomes bigger the faintest the volume-limited samples is, even though we decrease the scatter value to values close to zero. 
    \item We have also checked there is generally good agreement in the power spectrum monopole and quadrupole between the Uchuu lightcones and the DESI DR2 sample for BGS with $M_r < -20.0$ and LRG in the redshift bin \(0.6 < z < 0.8\), although the interpretation is complicated by the slightly different window functions.
    \item The linear bias factors were measured for both tracers from the DESI DR2 sample and compared to predictions based on the \textsc{Uchuu} lightcones in the Planck cosmology. We measured the bias as a function of absolute magnitude threshold for the various BGS volume-limited samples and as a function of the stellar mass for LRGs. For both tracers, we obtain that \textsc{Uchuu} bias is slightly larger than that obtained for DESI LRG Y3.
\end{itemize}

The results of this study are instrumental in refining key aspects of cosmological modelling, particularly in the generation of simulated lightcones and the development of more robust galaxy-halo connection techniques. These advancements will play a crucial role in the future of the DESI survey, enabling more accurate full-shape analyses of the Baryon Acoustic Oscillations (BAO) to constrain cosmological parameters. In particular, the mocks generated, especially the BGS catalog, will be invaluable for studies of large-scale structure, including the identification of voids. By enhancing our understanding of clustering signals, these findings will allow for a more precise interpretation of DESI’s large-scale structure observations, leading to better constraints on cosmological models and advancing our knowledge of the universe’s expansion and matter distribution.

A thorough analysis of the DESI DR2 Survey data allows us to assess and enhance the accuracy of methodologies used in cosmological studies, ultimately increasing the reliability of the final survey’s results. Furthermore, insights gained from the current DESI dataset provide a foundation for optimizing future observational strategies and theoretical models. These advancements are essential for maximizing the scientific impact of the full DESI survey, ensuring that it delivers the most accurate constraints on the fundamental properties of the universe.

\section{Data Availability}
The DR1 \textsc{Uchuu}-LRG and \textsc{Uchuu}-BGS lightcones will be made public at \url{https://data.desi.lbl.gov/doc/vac/}. DR2 lightcones will be made public with the DESI DR2. The data points corresponding to the figures from this paper will be available
in a Zenodo repository.

\acknowledgments

    EFG acknowledges financial support from the Severo Ochoa grant CEX2021-001131-S funded by MCIN/AEI/ 10.13039/501100011033.
    
    EFG and FP thank Instituto de Astrofisica de Andalucia (IAA-CSIC), Centro de Supercomputacion de Galicia (CESGA) and the Spanish academic and research network (RedIRIS) in Spain for hosting Uchuu DR1, DR2 and DR3 in the Skies $\&$ Universes site for cosmological simulations. The Uchuu simulations were carried out on Aterui II supercomputer at Center for Computational Astrophysics, CfCA, of National Astronomical Observatory of Japan, and the K computer at the RIKEN Advanced Institute for Computational Science. The Uchuu Data Releases efforts have made use of the skun$@$IAA$\_$RedIRIS and skun6$@$IAA computer facilities managed by the IAA-CSIC in Spain (MICINN EU-Feder grant EQC2018-004366-P).
    
    EFG and FP acknowledge support from the Spanish MICINN funding grant PGC2018-101931-B-I00. 

    TI has been supported by IAAR Research Support Program in Chiba University Japan, MEXT/JSPS KAKENHI (Grant Number JP19KK0344 and JP25H00662), MEXT as “Program for Promoting Researches on the Supercomputer Fugaku” (JPMXP1020230406), and JICFuS.

    This material is based upon work supported by the U.S. Department of Energy (DOE), Office of Science, Office of High-Energy Physics, under Contract No. DE–AC02–05CH11231, and by the National Energy Research Scientific Computing Center, a DOE Office of Science User Facility under the same contract. Additional support for DESI was provided by the U.S. National Science Foundation (NSF), Division of Astronomical Sciences under Contract No. AST-0950945 to the NSF’s National Optical-Infrared Astronomy Research Laboratory; the Science and Technology Facilities Council of the United Kingdom; the Gordon and Betty Moore Foundation; the Heising-Simons Foundation; the French Alternative Energies and Atomic Energy Commission (CEA); the National Council of Humanities, Science and Technology of Mexico (CONAHCYT); the Ministry of Science, Innovation and Universities of Spain (MICIU/AEI/10.13039/501100011033), and by the DESI Member Institutions: \url{https://www.desi.lbl.gov/collaborating-institutions}. Any opinions, findings, and conclusions or recommendations expressed in this material are those of the author(s) and do not necessarily reflect the views of the U. S. National Science Foundation, the U. S. Department of Energy, or any of the listed funding agencies.

The authors are honored to be permitted to conduct scientific research on I'oligam Du'ag (Kitt Peak), a mountain with particular significance to the Tohono O’odham Nation.



\bibliographystyle{mod-apsrev4-2} 
\bibliography{references}


\appendix

\section{Linear regime of the TPCF}\label{linear_appendix}
\begin{figure*}
    \centering
    \includegraphics[width=0.49\linewidth]{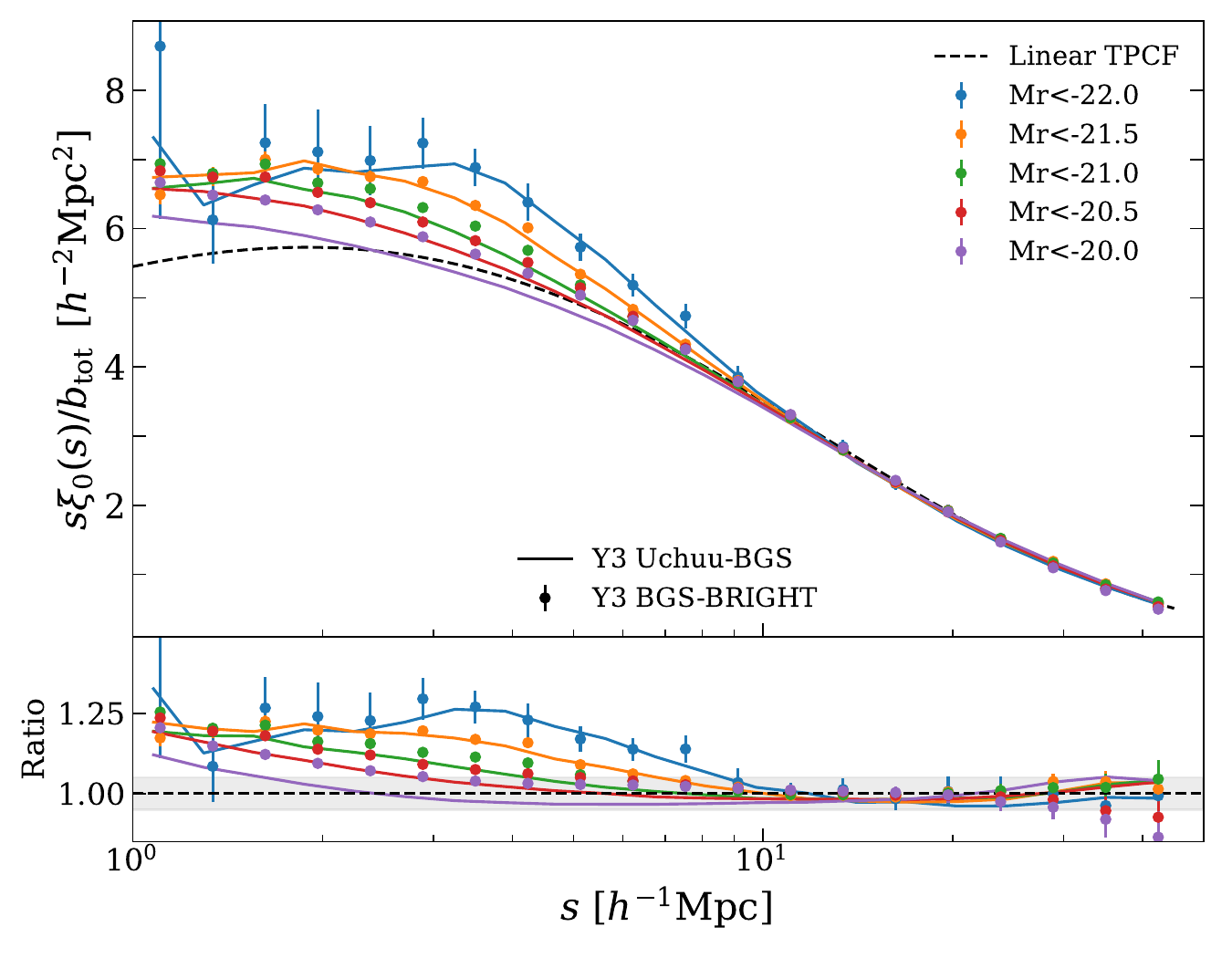}
    \includegraphics[width=0.49\linewidth]{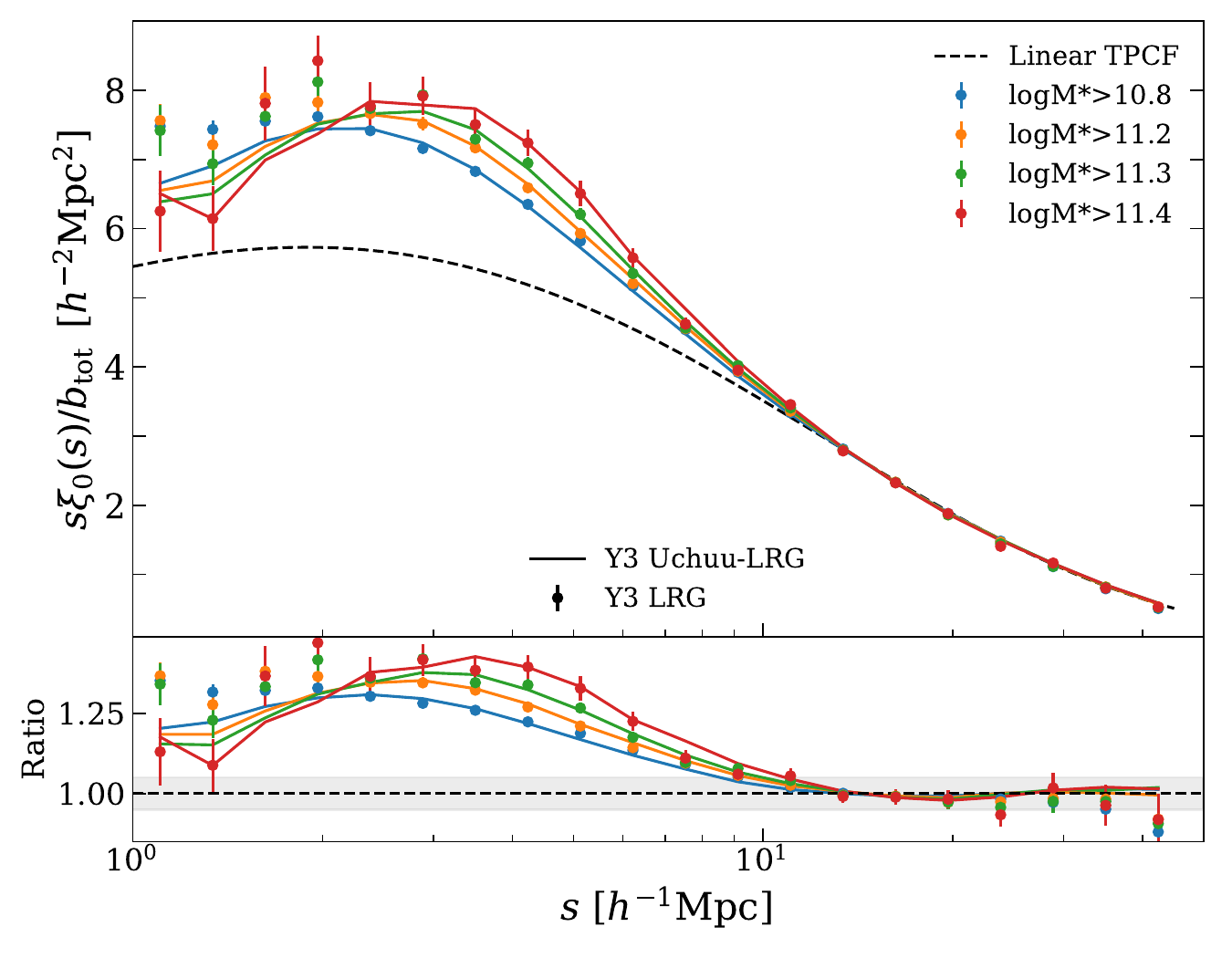}
    \caption{The left panel shows the comparison of the monopole of the linear Two-Point Correlation Function (TPCF) with the bias, $b$, measured at redshift $z = 0$, for different BGS volume-limited samples. The linear TPCF from the Uchuu simulation (dashed black line), also measured at redshift $z = 0$, is shown for reference. The right panel displays the same comparison, but for different LRG samples with various stellar mass cuts. The bottom panels show the ratio of the data points to the lines, relative to the linear TPCF. The shaded area represents the 5$\%$ limits. $b_{\rm tot}$ is equal to $b^{2}\left(1+\frac{2}{3}\beta+\frac{1}{5}\beta^{2} \right)$}
    \label{linear_tpcf}
\end{figure*}

In this section, we show that the TPCF is linear at separation scales between $10<s<40$ $h^{-1}$Mpc, which is the separation range considered in order to measure the large-scale bias in Section \ref{hod_bias}. This can be seen in Figure \ref{linear_tpcf}, where the linear Two-Point Correlation Function (TPCF) using the bias measurements derived in the aforementioned section for each BGS volume-limited sample and LRG sample with different stellar mass cuts. The left panel displays the linear TPCF at redshift $z=0$ for the data (represented by points) and mocks (depicted by lines) for BGS-BRIGHT targets, while the right panel shows the corresponding data for LRGs. Additionally, the linear TPCF calculated from Uchuu's matter power spectrum is shown as a dark dashed line. The bottom panels of both plots present the ratio between the data and mock linear TPCF. The shaded gray area represents the 5$\%$ limits. From these panels, it is evident that the agreement between the data and mock linear TPCF with that from Uchuu is excellent at scales between $10 < s < 20$ $h^{-1}$Mpc (the ratio is below a 5$\%$), and remains strong even at larger scales. For the calculation of the large-scale bias, we assumed that the TPCF was linear within the range $10 < s < 20$ $h^{-1}$Mpc, and this result confirms the validity of that assumption.

\section{Modelling of the colour distribution}\label{modelling_colour}

\begin{figure*}
    \centering
    \includegraphics[width=\linewidth]{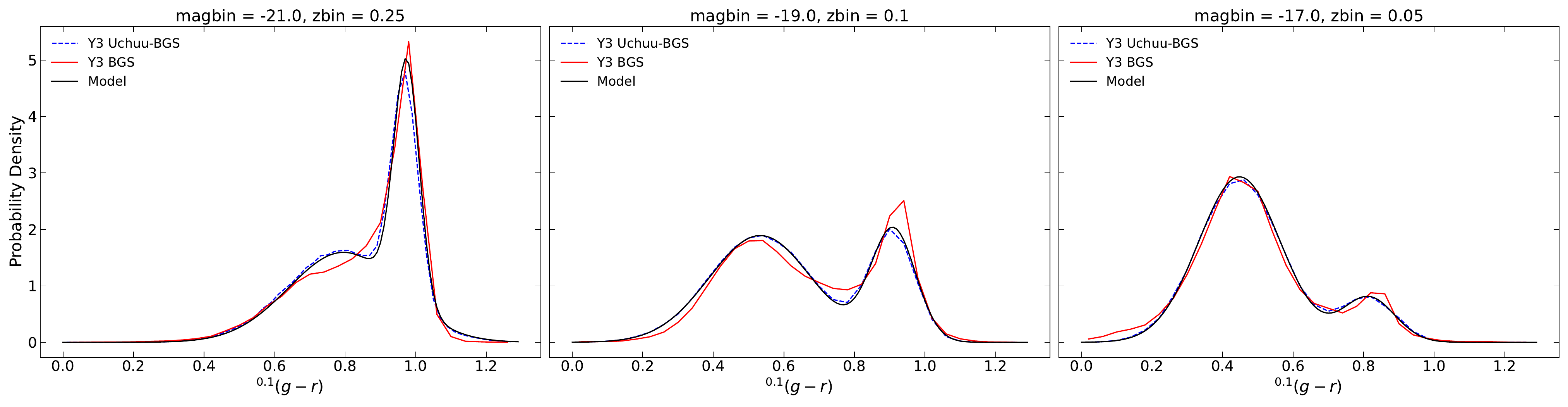}
    \caption{Comparison of the probability density of the colour distribution for Y3 BGS-BRIGHT (red line), Y3 Uchuu-BGS mock (blue line) and the model adopted for assigning a colour to each galaxy (black line). Each panel shows the result for different absolute magnitude bins (magbin) and redshift bins (zbin). The bin width of magbin and zbin is 0.2 and 0.02, respectively. }
    \label{color_comparison}
\end{figure*}

In this appendix, we present a comparison of the probability density functions for the colour distribution of the mock data, observational data, and the adopted model. This comparison is illustrated in Figure \ref{color_comparison}, where each panel corresponds to different bins of magnitude and redshift. By construction, the agreement between the mock data and the model is excellent. However, the model does not match the observational data as well, particularly for the faintest samples in the tails of the distribution (i.e., the reddest and bluest galaxies). This discrepancy arises because the model assumes a double Gaussian fit to the data, which is not always a suitable approximation for certain magnitude and redshift bins. Addressing this limitation will be crucial for improving the next generation of mock catalogues.

\end{document}

%% file: DESI-2024-0514_Mar24.tex

\author[1]{E.~Fernández-García\orcidlink{0009-0006-2125-9590}}
\affiliation[1]{Instituto de Astrof\'{i}sica de Andaluc\'{i}a (CSIC), Glorieta de la Astronom\'{i}a, s/n, E-18008 Granada, Spain}

\author[1]{F.~Prada\orcidlink{0000-0001-7145-8674}}

\author[2]{A.~Smith\orcidlink{0000-0002-3712-6892}}
\affiliation[2]{Institute for Computational Cosmology, Department of Physics, Durham University, South Road, Durham DH1 3LE, UK}

\author[3]{J.~DeRose\orcidlink{0000-0002-0728-0960}}
\affiliation[3]{Physics Department, Brookhaven National Laboratory, Upton, NY 11973, USA
}

\author[4,5,6]{A.~J.~Ross\orcidlink{0000-0002-7522-9083}}
\affiliation[4]{Center for Cosmology and AstroParticle Physics, The Ohio State University, 191 West Woodruff Avenue, Columbus, OH 43210, USA}
\affiliation[5]{Department of Astronomy, The Ohio State University, 4055 McPherson Laboratory, 140 W 18th Avenue, Columbus, OH 43210, USA}
\affiliation[6]{The Ohio State University, Columbus, 43210 OH, USA}

\author[7]{S.~Bailey\orcidlink{0000-0003-4162-6619}}
\affiliation[7]{Lawrence Berkeley National Laboratory, 1 Cyclotron Road, Berkeley, CA 94720, USA}

\author[8]{M. S.~Wang\orcidlink{0000-0002-2652-4043}}
\affiliation[8]{Institute for Astronomy, University of Edinburgh, Royal Observatory, Blackford Hill, Edinburgh EH9 3HJ, UK}

\author[9]{Z.~Ding\orcidlink{0000-0002-3369-3718}}
\affiliation[9]{University of Chinese Academy of Sciences, Nanjing 211135, People's Republic of China.}

\author[8]{C.~Guandalin\orcidlink{0000-0002-3369-3718}}

\author[10]{C.~Lamman\orcidlink{0000-0002-6731-9329}}
\affiliation{Center for Astrophysics $|$ Harvard \& Smithsonian, 60 Garden Street, Cambridge, MA 02138, USA}

\author[11]{R.~Vaisakh\orcidlink{0009-0001-2732-8431}}
\affiliation[11]{Department of Physics, Southern Methodist University, 3215 Daniel Avenue, Dallas, TX 75275, USA}

\author[11]{R.~Kehoe}

\author[12]{J.~Lasker\orcidlink{0000-0003-2999-4873}}
\affiliation[12]{Astrophysics \& Space Institute, Schmidt Sciences, New York, NY 10011, USA}

\author[13]{T.~Ishiyama\orcidlink{0000-0002-5316-9171}}
\affiliation[13]{Digital Transformation Enhancement Council, Chiba University, 1-33, Yayoi-cho, Inage-ku, Chiba, 263-8522, Japan}

\author[2]{S. M. ~Moore\orcidlink{0000-0002-5954-7903}}

\author[2]{S.~Cole\orcidlink{0000-0002-5954-7903}}

\author[14]{M.~Siudek\orcidlink{0000-0002-6186-5476}}
\affiliation[14]{Institute of Space Sciences, ICE-CSIC, Campus UAB, Carrer de Can Magrans s/n, 08913 Bellaterra, Barcelona, Spain
}

\author[11]{A.~Amalbert}

\author[15]{A.~Salcedo}
\affiliation[15]{Steward Observatory, University of Arizona, 933 N. Cherry Avenue, Tucson, AZ 85721, USA}

\author[16]{A.~Hearin}
\affiliation[16]{Argonne National Laboratory, High-Energy Physics Division, 9700 S. Cass Avenue, Argonne, IL 60439, USA}

\author[17]{B.~Joachimi}
\affiliation[17]{Department of Physics \& Astronomy, University College London, Gower Street, London, WC1E 6BT, UK}

\author[18]{A.~Rocher\orcidlink{0000-0003-4349-6424}}
\affiliation[18]{Institute of Physics, Laboratory of Astrophysics, \'{E}cole Polytechnique F\'{e}d\'{e}rale de Lausanne (EPFL), Observatoire de Sauverny, Chemin Pegasi 51, CH-1290 Versoix, Switzerland}

\author[19]{S.~Saito\orcidlink{0000-0002-6186-5476}}
\affiliation[19]{Institute for Multi-messenger Astrophysics and Cosmology, Department of Physics, Missouri University of Science and Technology, 1315 N Pine St, Rolla, MO 65409 U.S.A.
}

\author[20, 21, 22]{A.~Krolewski}
\affiliation[20]{Department of Physics and Astronomy, University of Waterloo, 200 University Ave W, Waterloo, ON N2L 3G1, Canada}
\affiliation[21]{Perimeter Institute for Theoretical Physics, 31 Caroline St. North, Waterloo, ON N2L 2Y5, Canada}
\affiliation[22]{Waterloo Centre for Astrophysics, University of Waterloo, 200 University Ave W, Waterloo, ON N2L 3G1, Canada}

\author[23]{Z.~Slepian}
\affiliation[23]{Department of Astronomy, University of Florida, 211 Bryant Space Science Center, Gainesville, FL 32611, USA}

\author[24]{Q.~Li\orcidlink{0000-0003-3616-6486}}
\affiliation[24]{Department of Physics and Astronomy, The University of Utah, 115 South 1400 East, Salt Lake City, UT 84112, USA
}

\author[24]{K.~S.~Dawson\orcidlink{0000-0002-0553-3805}}

\author[25]{E.~Jullo\orcidlink{0000-0002-9253-053X}}
\affiliation{[25]Aix-Marseille Univ., CNRS, CNES, LAM, Marseille, France}

\author[7]{J.~Aguilar}

\author[26]{S.~Ahlen\orcidlink{0000-0001-6098-7247}}
\affiliation[26]{Physics Dept., Boston University, 590 Commonwealth Avenue, Boston, MA 02215, USA}

\author[27,28]{D.~Bianchi\orcidlink{0000-0001-9712-0006}}
\affiliation[27]{Dipartimento di Fisica ``Aldo Pontremoli'', Universit\`a degli Studi di Milano, Via Celoria 16, I-20133 Milano, Italy}
\affiliation[28]{INAF-Osservatorio Astronomico di Brera, Via Brera 28, 20122 Milano, Italy}

\author[17]{D.~Brooks}

\author[7]{T.~Claybaugh}

\author[29]{A.~de la Macorra\orcidlink{0000-0002-1769-1640}}
\affiliation[29]{Instituto de F\'{\i}sica, Universidad Nacional Aut\'{o}noma de M\'{e}xico,  Circuito de la Investigaci\'{o}n Cient\'{\i}fica, Ciudad Universitaria, Cd. de M\'{e}xico  C.~P.~04510,  M\'{e}xico}

\author[17]{P.~Doel}

\author[30, 7]{S.~Ferraro\orcidlink{0000-0003-4992-7854}}
\affiliation[30]{University of California, Berkeley, 110 Sproul Hall \#5800 Berkeley, CA 94720, USA}

\author[31]{A.~Font-Ribera\orcidlink{0000-0002-3033-7312}}
\affiliation[31]{Institut de F\'{i}sica d’Altes Energies (IFAE), The Barcelona Institute of Science and Technology, Edifici Cn, Campus UAB, 08193, Bellaterra (Barcelona), Spain
}

\author[32,33]{J.~E.~Forero-Romero\orcidlink{0000-0002-2890-3725}}
\affiliation[32]{Departamento de F\'isica, Universidad de los Andes, Cra. 1 No. 18A-10, Edificio Ip, CP 111711, Bogot\'a, Colombia}
\affiliation[33]{Observatorio Astron\'omico, Universidad de los Andes, Cra. 1 No. 18A-10, Edificio H, CP 111711 Bogot\'a, Colombia}

\author[7]{S.~Gontcho A Gontcho\orcidlink{0000-0003-3142-233X}}

\author[34]{G.~Gutierrez}
\affiliation[34]{Fermi National Accelerator Laboratory, PO Box 500, Batavia, IL 60510, USA}

\author[4,35,6]{K.~Honscheid\orcidlink{0000-0002-6550-2023}}
\affiliation[35]{Department of Physics, The Ohio State University, 191 West Woodruff Avenue, Columbus, OH 43210, USA}

\author[36]{M.~Ishak\orcidlink{0000-0002-6024-466X}}
\affiliation[36]{Department of Physics, The University of Texas at Dallas, 800 W. Campbell Rd., Richardson, TX 75080, USA}

\author[37]{R.~Joyce\orcidlink{0000-0003-0201-5241}}
\affiliation[37]{NSF NOIRLab, 950 N. Cherry Ave., Tucson, AZ 85719, USA}

\author[37]{S.~Juneau\orcidlink{0000-0002-0000-2394}}

\author[38]{D.~Kirkby\orcidlink{0000-0002-8828-5463}}
\affiliation[38]{Department of Physics and Astronomy, University of California, Irvine, 92697, USA}

\author[7]{T.~Kisner\orcidlink{0000-0003-3510-7134}}

\author[7]{A.~Kremin\orcidlink{0000-0001-6356-7424}}

\author[17]{O.~Lahav}

\author[7]{A.~Lambert}

\author[7]{M.~Landriau\orcidlink{0000-0003-1838-8528}}

\author[7]{M.~E.~Levi\orcidlink{0000-0003-1887-1018}}

\author[39,40]{M.~Manera\orcidlink{0000-0003-4962-8934}}
\affiliation[39]{Departament de F\'{i}sica, Serra H\'{u}nter, Universitat Aut\`{o}noma de Barcelona, 08193 Bellaterra (Barcelona), Spain}
\affiliation[40]{Institut de F\'{i}sica d’Altes Energies (IFAE), The Barcelona Institute of Science and Technology, Edifici Cn, Campus UAB, 08193, Bellaterra (Barcelona), Spain}

\author[40,41]{R.~Miquel}
\affiliation[40]{Institut de F\'{i}sica d’Altes Energies (IFAE), The Barcelona Institute of Science and Technology, Edifici Cn, Campus UAB, 08193, Bellaterra (Barcelona), Spain}
\affiliation[41]{Instituci\'{o} Catalana de Recerca i Estudis Avan\c{c}ats, Passeig de Llu\'{\i}s Companys, 23, 08010 Barcelona, Spain}

\author[42]{J.~Moustakas\orcidlink{0000-0002-2733-4559}}
\affiliation[42]{Department of Physics and Astronomy, Siena College, 515 Loudon Road, Loudonville, NY 12211, USA}

\author[43]{S.~Nadathur\orcidlink{0000-0001-9070-3102}}
\affiliation[43]{Institute of Cosmology and Gravitation, University of Portsmouth, Dennis Sciama Building, Portsmouth, PO1 3FX, UK}

\author[44,45]{W.~J.~Percival\orcidlink{0000-0002-0644-5727}}
\affiliation[44]{Perimeter Institute for Theoretical Physics, 31 Caroline St. North, Waterloo, ON N2L 2Y5, Canada
}
\affiliation[45]{Waterloo Centre for Astrophysics, University of Waterloo, 200 University Ave W, Waterloo, ON N2L 3G1, Canada
}

\author[46]{I.~P\'erez-R\`afols\orcidlink{0000-0001-6979-0125}}
\affiliation[46]{Departament de F\'isica, EEBE, Universitat Polit\`ecnica de Catalunya, c/Eduard Maristany 10, 08930 Barcelona, Spain}

\author[47]{G.~Rossi}
\affiliation[47]{Department of Physics and Astronomy, Sejong University, 209 Neungdong-ro, Gwangjin-gu, Seoul 05006, Republic of Korea}

\author[48]{E.~Sanchez\orcidlink{0000-0002-9646-8198}}
\affiliation[48]{CIEMAT, Avenida Complutense 40, E-28040 Madrid, Spain}

\author[7]{D.~Schlegel}

\author[49]{H.~Seo\orcidlink{0000-0002-6588-3508}}
\affiliation[49]{Department of Physics \& Astronomy, Ohio University, 139 University Terrace, Athens, OH 45701, USA}

\author[7]{J.~Silber\orcidlink{0000-0002-3461-0320}}

\author[37]{D.~Sprayberry}

\author[50]{G.~Tarl\'{e}\orcidlink{0000-0003-1704-0781}}
\affiliation[50]{Department of Physics, University of Michigan, 450 Church Street, Ann Arbor, MI 48109, USA}

\author[37]{B.~A.~Weaver}

\author[51]{P.~Zarrouk\orcidlink{0000-0002-7305-9578}}
\affiliation[51]{Sorbonne Universit\'{e}, CNRS/IN2P3, Laboratoire de Physique Nucl\'{e}aire et de Hautes Energies (LPNHE), FR-75005 Paris, France}

\author[7]{R.~Zhou\orcidlink{0000-0001-5381-4372}}
